# A robust and tunable Luttinger liquid in correlated edge of transition-metal second-order topological insulator Ta$_2$Pd$_3$Te$_5$

Anqi Wang[1,2,†], Yupeng Li[1,†], Guang Yang[1,†], Dayu Yan[1,†], Yuan Huang[3], Zhaopeng Guo[1], Jiacheng Gao[1,2], Jierui Huang[1,2], Qiaochu Zeng[1], Degui Qian[1], Hao Wang[1], Xingchen Guo[1,2], Fanqi Meng[1], Qinghua Zhang[1,4], Lin Gu[1,2,5], Xingjiang Zhou[1,2,5], Guangtong Liu[1,5], Fanming Qu[1,2,5], Tian Qian[1,5], Youguo Shi[1,2,5*], Zhijun Wang[1,2*], Li Lu[1,2,5*], Jie Shen[1,5*]

[1]Beijing National Laboratory for Condensed Matter Physics, Institute of Physics, Chinese Academy of Sciences, Beijing 100190, China

[2]School of Physical Sciences, University of Chinese Academy of Sciences, Beijing 100049, China

[3]Advanced Research Institute of Multidisciplinary Science, Beijing Institute of Technology, Beijing 100081, China

[4]Yangtze River Delta Physics Research Center Co. Ltd, Liyang 213300, China

[5]Songshan Lake Materials Laboratory, Dongguan 523808, China

†These authors contributed equally to this work

*Corresponding author. Email: ygshi@iphy.ac.cn (Y.-G.S.), wzj@iphy.ac.cn (Z.-J.W.), lilu@iphy.ac.cn (L.L.), shenjie@iphy.ac.cn (J.S.)

**Abstract**

**The interplay between topology and interaction always plays an important role in condensed matter physics and induces many exotic quantum phases, while rare transition metal layered material (TMLM) has been proved to possess both. Here we report a TMLM Ta$_2$Pd$_3$Te$_5$ has the two-dimensional second-order topology (also a quadrupole topological insulator) with correlated edge states - Luttinger liquid. It is ascribed to the unconventional nature of the mismatch between charge- and atomic- centers induced by a remarkable double-band inversion. This one-dimensional protected edge state preserves the Luttinger liquid behavior with robustness and universality in scale from micro- to macro- size, leading to a significant anisotropic electrical transport through two-dimensional sides of bulk materials. Moreover, the bulk gap can be modulated by the thickness, resulting in an extensive-range phase diagram for Luttinger liquid. These provide an attractive model to study the interaction and quantum phases in correlated topological systems.**

**Main Text:**

**Introduction**

Whereas topological systems without consideration of correlation have been studied widely in the last decades, the exotic quantum phases derived from the interaction between electrons and topological state have recently attracted great attention[1], such as quantum spin liquids[2] and topological Mott insulators[3] which possess one-dimensional (1D) bosonic mode with spin-charge separation, strongly correlated Chern insulators in flat bands of magic-angle twisted bilayer graphene[4,5] and topological superconductors for topological quantum computation[6,7]. The aforementioned correlated 1D bosonic mode is described as Tomonoga-Luttinger liquid (simplified as Luttinger liquid, LL), and characterized by a power-law vanishing to anomaly zero density-of-state at the Fermi level. This power-law behavior in electrical transport measurements has been harnessed to experimentally test LL in the edge states of quantum spin Hall insulator candidates[8-10] and quantum Hall systems[11-16]. Thus, LL quantum phase provides a powerful tool to analyze the correlated 1D edge state emerging from topological order. However, in these systems, the LL relies on the microscopic details and is lack of experimental evidence for its universality and robustness, let alone extending to macro-scale, which is supposed to be a signature for topological order. Moreover, this vulnerability leads to the absence of the quantitative test of LL, e.g., its exponent factor as the Coulomb interaction strength, in the edge state of topological system, though a large amount of theories have discussed the possibility of taking into account different interactions[17-21].

On the other hand, unlike a two-dimensional (2D) topological insulator with gapless edge states, a 2D second-order topological insulator (SOTI, or a quadrupole topological insulator) is supposed to have gapped edge states and in-gap corner states (in the presence of chiral symmetry)[22]. Although many 2D SOTI candidates have been proposed in literature[23-27], rare transition-metal compound has been predicted to be a 2D SOTI. After considering many-body interactions, superconductivity, excitonic condensation and Luttinger liquid can be expected in bulk and edge states of the transition-metal SOTI.

Here, we have observed the LL inhabiting the edge states, which are as the remnants of the 2D second-order topology, of a TMLM $Ta_2Pd_3Te_5$. Through the first-principles calculations, we demonstrate that this topological state originates from a remarkable double-band inversion and results in an unconventional nature with an essential band representation A"@4c[28]. Then we have performed the electrical transport measurements of tens of $Ta_2Pd_3Te_5$ bulk and thin-film devices, and detected the coexistence of the bulk insulating gap, the in-gap edge conducting state and the edge gap as the signature of the 2D SOTI. Interestingly, the edge state displays a universal scaling of power-law relation with a zero-energy anomaly as a function of temperature ($T$) and energy (signed as the voltage bias here), which is considered as the critical evidence of LL. The weak van der Waals (vdW) interaction between layers allows this LL to be reproducible in the edge channels of all the devices from atomic- to macro- scale, with a conductance proportional to the edge geometry. That is the 2D side facets of macroscopic bulk also show 1D LL anisotropic electrical transport. This reveals for the first time the evidence of a scale connection from atomic- to macro- LL, and supports this LL quantum state with universality and robustness irrespective of the microscopic details. In addition, the coexistence of the three states induces a tunability of the LL power-law exponent to represent the strength of Coulomb interaction, as well as a reversible LL-Fermi liquid (FL) transition and a full phase diagram. All these demonstrate a robust and universal LL with the many precise-controlled parameters, as an effective tool to characterize the interaction and the quantum phase transition in the correlated topological system.

**Crystal structure**

$Ta_2Pd_3Te_5$ is a vdW material with quasi-one-dimensional (Q1D) $Ta_2Te_5$ chains and crystallizes in needle-like single crystals along the *b*-direction (see the crystal structure and the optical image of the typical samples in Fig. 1a and ref. [29]). The single crystal is synthesized with an orthorhombic structure *Pnma* (space group No.62)[29]. The Q1D characteristic and layered structure are confirmed by high-resolution scanning transmission electron microscope (STEM) images of $Ta_2Pd_3Te_5$ single crystal along [010] and [001] projections, revealing a thickness of ~0.7 nm for monolayer (Fig. 1b and Supplementary Fig. 1, b and c). Because of the strong anisotropic bonding energy[29], this material is easy to mechanically exfoliate into large and thin films with flat and uniform edges along the Q1D chain (Supplementary Fig. 1a). Importantly, the Fermi level naturally locates in the bottom of conduction bands (Fig. 1c and Supplementary Fig. 1d), enabling a high-efficient modulation by the electrostatic gate. The coupling between vdW layers is weak, proved by the angle resolved photoemission spectroscopy (APRES) measurement (Fig. 1d and Supplementary Fig. 1, e and f).

**Theoretical calculation**

Recently, the Perdew-Burke-Ernzerhof (PBE) calculations with spin-orbit coupling (SOC) predict that its monolayer is a quantum spin Hall state with a tiny SOC gap and the bulk topological crystalline insulator without a global band gap (ref. [29]). However, both of the scanning tunneling microscopy (STM) (ref. [30]) and APRES (Fig. 1c) measurements detect a bulk gap of ~50 meV. As reported in ref. [29], the nearly zero-energy band gap is sensitive to the lattice constant. On the other hand, the experimental energy gap can be induced by many-body interactions in the Q1D structure, like electron-phonon interaction or electron-hole attractive interaction, as the excitonic state is thermodynamically most stable in a zero-gap semiconductor. Although the exact reason of the bulk gap opening needs further investigations, the double-band inversion happens clearly between the {Y2;GM1-,GM3-} and {Y1;GM1+,GM3+} bands, which are 0.3 eV (or higher) away from the Fermi level in the monolayer structure of SG59 (imposing inversion; Supplementary Table. 1 and 2)[28]. The schematic of the double band inversion in the system is given in Fig. 1e. Hence, in this manuscript, we focus on the modified band structure of the monolayer $Ta_2Pd_3Te_5$ (by adding small onsite energy to simulate the many-body gap opening) in Fig. 1f[31]. The band representation (BR) decomposition of the occupied bands shows that there is an essential BR at vacancies, suggesting the unconventional nature/obstructed atomic limit with second-order topology[31, 32]. A 2D quadrupole topological insulator with second-order topology possesses 1D gapped edge states and isolated corner states (Detailed information can be found in Ref. 28). In the computed (01)-edge spectrum of Fig 1g, the gapped edge states are clearly shown in the gap of the bulk continuum. The observed in-gap edge states as the remnants of the quadrupole topological insulator can exhibit 1D feature well, which is responsible for the observed LL behavior (will be analyzed later). In addition, since the layers of the bulk are vdW coupled (Fig. 1d and Supplementary Fig. 1, e and f), the 1D edge states can still survive on the side surfaces of the bulk.

**Three features in d$I$/d$V$ curves**

To investigate the electronic properties, we show the typical result for thin-film devices #A1,2 with thickness ~10 nm in Fig 2, with a back-gate voltage $V_{bg}$ tuning the chemical potential as illustrated in Fig. 2a. Figure 2b and c show the four-terminal measurement configurations and gate tunable d$I$/d$V$ for devices #A1,2. Once applying the voltage bias ($V_{DC}$), the differential conductance (d$I$/d$V$($V_{DC}$)) curve of #A1 at charge neutral point (CNP, $V_{bg}$ =-28V) shows distinguishable three regions (Fig. 2d, indicated by the purple-green-orange background). The first region at small $V_{DC}$ indicates a gap structure (highlighted by the purple background, also see the conductance valley plotted by the symmetric $V_{DC}$ in the inset of Fig. 2d), which closes around $T$ = 3.75 K; The second region at medium $V_{DC}$ reveals a power-law shape with an exponent of ~0.33 (highlighted by the green background); The third region at high $V_{DC}$ shows a new shape of enhanced conductance with increasing $V_{DC}$. Those regions are also repeated in other devices (Fig. 2e for device #A2 and Supplementary Fig. 12, l and m). Notably, for device #A2, the $T$-dependent d$I$/d$V$($T$) curve at CNP (purple curve in Fig. 2f) shows similar three regions, which are semiconductor behavior with an energy gap of ~27 meV from 300 to ~50 K, a power-law dependence with the exponential factor α of ~0.79 from ~30 K to 6 K and another semiconductor behavior with a smaller energy gap of ~ 0.6 meV at lower $T$. This is clearer in the log(d$I$/d$V$) v.s. 1/$T$ plot in the inset where the yellow dot dash lines are fitting curves of Arrhenius equation d$I$/d$V$($T$) ∝ exp(−Δ/2k$_B$$T$). Similar behavior is repeatable, e.g., device #A4 with a ~ 33 meV gap at high $T$ and a ~ 0.9 meV gap at low $T$ in Supplementary Fig. 2. Apparently, this three-region feature fits the theoretical result of band structure with the bulk gap $E_b$, edge conducting state and edge gap $E_e$ very well.

It should be noted that these devices are measured with four-terminal configuration with standard lock-in measurement and the phases $θ$ of the lock-in amplifier are also small enough. So, the very-low-$T$ behaviors do not come from the contact resistance which is not included in the measurement. In addition, the gap size we extracted from the Arrhenius equation is consistent with the thermal excitation energy of the temperature at which the gap behavior appeared. We also measured d$I$/d$V$($V_{DC}$) and d$I$/d$V$($T$) from CNP to positive back-gate voltage ($V_{bg}$) (Fig. 2, e and f). As the Fermi level moves up via applying positive $V_{bg}$ sketched in Fig. 2a, the three-region feature gradually transitions into that of two regions without $E_e$ in both d$I$/d$V$($V_{DC}$) and d$I$/d$V$($T$) curves, further consistent with the calculated band structure.

**Validation of edge state**

   To better clarify where the three structures respectively come from, we perform the nonlocal transport measurement based on Hall bar configuration[33]. For a classical planar Hall bar, when a local voltage $V_0$ is applied, the nonlocal response $V_{nl}$ at remote contacts will decay exponentially with distance, irrespective of $T$, chemical potential $\mu$ and so on. Whereas, the nonlocal voltages through edge channels don't rely on the distance[34]. Fig.3a is a typical example of device #A3 with local and nonlocal measurement configuration (see Supplementary Fig. 5 for nonlocal measurement of device #A2). For the nonlocal conductance in Fig 3, c and d, both of nonlocal ratios $V_{nl1}/V_0$ and $V_{nl2}/V_0$ at CNP increase with decreasing $T$ gradually between 50-120 K (see the orange region in insets of Fig. 3, c and d), which are due to the reduced contact resistance of local electrodes and still behave as the classical ohmic planar fluid (the dash lines in the inset are classical planar values obtained by the simulation in Supplementary Fig. 6). Afterwards, they grow drastically on cooling from 50 K to 9 K (the green region), reaching values far above the classical values and indicating the edge transport case. Particularly, the sum of these two nonlocal ratios is close to 1 around 6 K. In the same $T$ region (Fig. 3b), a power-law region appears in the local conductance fitted by the dash line (inset) and revealing it is the edge state. Interestingly below 4 K when the local conductance enters the $E_e$ (purple region), $V_{nl1}/V_0$ comes up to 1 and $V_{nl2}/V_0$ goes to zero, suggesting that the inhomogeneity in $E_e$ between different edge channels is enhanced. $V_{nl}/V_0(V_{bg})$ curves at low $T$ in Fig 3, c and d show nonlocal values become classic ohmic planar-like when Fermi level goes towards the conduction band at positive $V_{bg}$, excluding these unusual nonlocal values are due to planar fluid with extreme-disordered cases, e.g., with cracks or being broken into pieces. Therefore, by comparing the local and nonlocal measurements at different $T$ and $V_{bg}$, we deduce both of the power-law feature and the small gap $E_e$ root in the edge channels. This also excludes $E_e$ from any bulk insulating state, e.g., Mott insulator or Anderson insulator, in particular with the phenomena of the greatly tunable $T$ range at samples with varying thickness (will be analyzed more later in section 'Luttinger liquid phase diagram').

   More evidence to prove this edge state is that while both APRES (Fig. 1c) and STM (ref. [30]) clearly detect a large gap up to ~ 50 meV on the top surface of $Ta_2Pd_3Te_5$ single crystal, the transport measurements still show large conductance around CNP on thin films exfoliated from

the same bulk sample. Some of the conductance even reach a value as high as $2G_0$ ($G_0$ is the quantized conductance $2e^2/h$). STM measurement also reveals a distinct local conducting edge state in the bulk energy gap[30]. Additionally, the similar edge conductivity in power-law region for samples with various thickness further confirms the existence of robust edge states from atomic- to macroscopic- scale and will be analyzed in detail in section 'Robust and universal Luttinger liquid from micro- to macro- size'. Moreover, the finite conductance around CNP, compared with the close-to-zero conductance when grounding the edge, also supports the existence of edge states (Supplementary Fig. 7). In conclusion, all the local and nonlocal measurements prove the existence of bulk state, edge conducting state and correlated edge gap, namely the second-order topology in this material.

**Tunable Luttinger liquid in edge states**

Now let's discuss the power-law relation of the edge state. Such behavior in 1D quantum system is a typical signature of LL, in particular with a universal scaling. We use device #A4 as a typical example whose behavior is similar to device #A2. Fig. 4a shows in $dI/dV(V_{DC})$ curves for device #A4 at $V_{bg} = 0$ V, $E_e$ disappears and power-law relation dominates at higher $T$ (other systematic data of device #A4 are shown in Supplementary Fig. 2). For those $dI/dV(V_{DC})$ curves without $E_e$ behavior (From $T = 12.9$ K to $39.8$ K), in Fig. 4b we scaled the conductance $(dI/dV)/T^{0.69}$ with exponent of ~0.69, which is extracted from $dI/dV(T)$ curve in Supplementary Fig. 2b, as $y$-axis and $eV_{DC}/k_BT$ ($k_B$ is the Boltzmann constant) as $x$-axis. By this, all the curves converge together with a zero-energy anomaly, confirming the universal power-law correlation as well as LL behavior (similar behavior of device #A2 is shown in Supplementary Fig. 4 and more examples are in Supplementary Fig. 2, 3, 12). We note that for $dI/dV(V_{DC})$ curves at the lowest temperature (e.g. $T < 12.9$ K in Fig. 4a), the low-bias regions exhibit edge gap behavior due to the 2D SOTI nature which limits the power-law and scaling ranges, so these low-temperature $dI/dV(V_{DC})$ curves shouldn't obey the scaling behavior physically.

Furthermore, we linearly tune the power-law exponent α from ~ 0.7 at CNP to below 0.1 at positive $V_{bg}$ (Fig. 2e and 4c, here α = 0 means FL). This reveals a tunability of the Coulomb interaction strength of 1D edge state, as well as the transition from LL of correlated 1D edge state to FL of extended high-dimensional bulk state along with the rising of Fermi level. The

exponent α is correlated with the strength of the Coulomb interaction and has been harnessed to test different types of interaction, e.g., Kondo effect in the helical edge state[19], filling in the fractional quantum Hall[17]. The similar slope of tunability by gate voltage (also indicating the electron carrier density) in Fig 4c from three devices might indicate a constant intrinsic interaction and provide a controllable approach to study the interaction strength and the quantum phase transition. Usually, one can use the exponent α to estimate the Luttinger liquid parameter $K$, which is an intrinsic parameter describing the electron-electron interactions in the system. But the numerical relationship between α and $K$ intricately depends on the details of the system, such as the type/strength of tunnel barriers[35-37], edge state of fractional quantum Hall effect[12, 38]/quantum spin Hall effect[8, 19], spin-orbit coupling[37], etc. The relation in our system is not clear yet, so we could not get the accurate calculation of $K$ here and further theoretical modelling is needed.

**Robust and universal Luttinger liquid from micro- to macro- size**

Here, we find there are two types of devices, among which we name the devices with distinguishable three regions as type A and the devices without $E_e$ as type B. Because the edge conducting state is dominating at low energy, device B owns straight power-law traces with the absence of edge gap for both d$I$/d$V$($T$) and d$I$/d$V$($V_{DC}$) curves, as well as large values of conductance, at small $T$ and $V_{DC}$ around CNP (Fig. 5a-c for device #B1 for example), similar to the curves at positive $V_{bg}$ for device A (see the green curves in Fig. 2, e and f). These power-law traces could also be scaled (Fig. 5c). Prominently, different sections of the same device could be scaled with the similar exponent (Supplementary Fig. 8). All these further confirm the intrinsic LL property of the edge conducting state, no matter w/o $E_e$. Interestingly, the exponent α from d$I$/d$V$($T$) curves is always larger than, sometimes nearly twice of, that from d$I$/d$V$($V_{DC}$) curve (Fig. 2 and 5, Supplementary Fig. 2-4 and 12-14), revealing both the inner LL-LL tunneling and FL-LL tunneling through electrodes are involved here[35, 36, 39, 40] - the exponents of LL-LL junctions are twice of FL-LL junctions. So, the LL-LL tunneling becomes dominant and the exponent of d$I$/d$V$($T$) is α(LL-LL) at low temperature due to the larger resistance of LL-LL junctions with larger exponents (with the assumption that both FL-LL and LL-LL junctions have similar resistance at room temperature). When we apply high bias, FL-LL

junctions become dominant and the exponent of d$I$/d$V$($V_{DC}$) is α(FL-LL) (FL-LL junctions become more resistive because their resistance decreases more slowly)[39, 40].

This edge LL is of great stability and reproducibility, because it shows up in almost all the devices and maintains as long as at least one year. Remarkably, bulk samples with macroscopic geometry of ~ 100 $\mu$m, which we name as type C, also exhibit the nice LL behavior with a universal power-law correlation down to 100 mK (Fig. 5d-f and Supplementary Fig. 14). Interestingly, both of the thick devices and bulk samples behave like type B without $E_e$. Thus, it gives us a hint that the reason for the absence of $E_e$ might be the larger fluctuation of chemical potential or slight band variation due to the coupling of increasing layers. It is important to emphasize that there is no substantial difference between type A and B devices (and type C bulk samples); the only difference is the edge gap size between them. Type B devices and type C bulk samples do not exhibit distinct edge gap behavior down to the lowest temperature (~ 1.5 K for type B devices and ~ 0.1 K for type C samples) even at CNP in the measurement. But we do not deny the possibility that they will exhibit edge gap behavior at a lower temperature (see Supplementary Fig. 8g for device #B4$_R$ as an example).

There is another point of interest, the kinks in the scaling curves of bulk samples with larger lead spacing are usually larger than that of thin films with smaller lead spacing (Fig. 5c and 5f as an example). In Luttinger liquid physics, the kink position in $eV/k_BT$ represents the number of the tunnel barriers in the 1D channel. We give a brief statistical result of this value for type B devices, type C bulk samples and Luttinger liquid region of type A devices in Supplementary Fig. 9a. It is clear that the number of tunnel barriers is positively correlated with the channel length.

Generally, Luttinger liquid behavior will be more credible if the power-law regions can extend to more than one decade. In our samples, at high temperature, thermal excitation will drive the sample exhibiting a bulk semiconducting behavior and at low temperature edge gap behavior will arise due to the 2D SOTI nature. The range of resistance change in power-law regions is smaller due to the small power exponents (usually < 1 for our devices/samples). Even so, the power-law range for almost all the type B devices/type C samples without edge gap and some of the type A devices can extend to more than one decade.

Apparent power-law dependence also shows up because of the environmental quantum fluctuation (EQF) or special region of variable range hopping (VRH)[10, 41–44]. For example, VRH from a disordered quasi-1D state due to inhomogeneity or folding lines in devices could show 'apparent' power-law behavior[44]. We can exclude the above-mentioned reasons as follows. Firstly, the high quality of our devices with PMMA protecting layers (see 'Devices fabrication' section in Methods for detail) can be verified by the optical images of the devices (inset of Fig. 2b,c, Fig. 3a, etc.). For a small number of devices with hBN as top capping layer (only 3 of the 22 devices - device #A6, Supplementary device #1 and #2), the optical images may exhibit contrast color (Supplementary Fig. 7) due to the unevenness of the top hBN on thick Ti/Au electrodes. These top capping layers are insulating and serve only as protecting layers, so they do not influence the transport behavior. This could also be confirmed by that we repeated robust power-law relation in various samples - with varying sizes from microscopic to macroscopic scale - at wide $T$ and $V_{DC}$ range other than specific samples with certain parameter range. Secondly, for VRH transport, the power exponent should become larger with increasing channel length[44]. As a contrast, in our result, the power exponent is similar for microscopic devices and bulk samples with hundreds of micrometers (Supplementary Fig. 9b), inconsistent with VRH transport. Lastly, we have further proved the robust existence of clean edge states, instead of random 1D disordered channels by the sawtooth-shaped SQUID (superconducting quantum interference device) pattern in the Al-$Ta_2Pd_3Te_5$-Al Josephson junction[45].

The feature that semiconductor/Luttinger behavior at high/low $T$ arising from bulk/edge will be more visual if we plot $T$-dependent curves of several devices B and bulk samples C together in bulk conductivity σ (Fig. 5g) or edge sheet conductivity $σ_{edge}$ (Fig. 5h). Because the values in high $T$ region converge into the same order of magnitude in the former (orange region in Fig. 5g) while that in low $T$ region in the latter (green region in Fig. 5h). As a contrast, the curves in top (bottom) surface sheet conductivity diverge in the entire $T$ region (Supplementary Fig. 10), which excludes the possibility of the existence of top/bottom surface states. Moreover, the starting resistance for LL ($R_{LL}$, defined as 2% deviation of the extension resistance from power-law behavior in LL region) increases almost linearly with the number of squares in edge, $□_{edge}$, for the devices with thickness of from ~ 4 nm to ~ 200 $\mu$m over five orders of magnitude (Fig.

5i). This is the first experimental observation of the scale connection between atomic- (equals to ~6 layers) and macro- LL, and confirms the LL quantum state with correlation on the length scale and a universal feature independent of the microscopic details[46, 47]. It also supports the edge state calculated for the monolayer is generic for bulk sample due to the weak inter-layer coupling as the sketch in the inset of Fig. 5i shows. Such an in-gap 1D LL in the 2D sides of bulk $Ta_2Pd_3Te_5$ might boost a potential application of this giant anisotropic electrical transport, as well as the anisotropic heating transport.

The temperature-dependent electrical resistance of the bulk $Ta_2Pd_3Te_5$ has also been measured in previous works and the results are consistent with ours[30,48]. Interestingly, as reported in Ref. 48, the superconducting behavior can be observed in Ti- or W-doped $Ta_2Pd_3Te_5$, indicating luxuriant physics phenomena hidden in $Ta_2Pd_3Te_5$ system.

**Luttinger liquid phase diagram**

We have mentioned the weak inter-layer coupling in Fig. 1d, but interestingly the bulk gap still varies with thickness. A plausible reason for the increasing bulk gap with decreasing thickness (summarized in Fig. 6a, calculated from the Arrhenius equation $\Delta = -2k_B(\partial \ln(dI/dV)/\partial T^{-1})$ by using the $dI/dV(T)$ curve) is that the insulating state is more robust in thinner devices due to the reduced screening effect to the Coulomb interaction. As the appearance of LL relies on both of the bulk gap and edge gap, it is easy to perform a quantitative study of LL phase diagram. Fig. 6b illustrates the summary of starting and ending temperature, $T_1$ and $T_2$, of LL from $dI/dV(T)$ curves of devices with different thicknesses and results in a clear thickness-dependence phase diagram of bulk semiconducting FL, edge LL and edge gap. The temperature range of LL is adjustable over three orders of magnitude from ~20 - 100 K in thin film devices to 0.1 - 7 K in bulk samples, exhibiting a large parameter space. We also find $T_1$, which indicates the energy size of $E_b$, varies for a large temperature range with thickness but is still correlated with $T_2$ (indicating the size of $E_e$, Supplementary Fig. 11). This reveals that $E_e$ originates from the intrinsic band structure instead of other insulating state.

**Summary**

As such, this LL has not only extended the LL families to the more general boundary, beyond helical and chiral edges, of high-dimensional systems, but also remarkably promotes the reproducibility and enlarges parameter space including geometry, temperature (*T*) and

chemical potential ($\mu$). We conclude this LL possesses universality and robustness, as well as a full phase diagram with extensive range, complete data set and strong tunability. Therefore, this model system would have wide-reaching implications to understand quantum phases towards more complex systems, with a combination of strong correlation and topology, which is lack of useful and practical method to analyze before.

## Methods

**Density functional theory calculations**

The first-principles calculations were performed within the framework of the density functional theory (DFT) using the projector augmented wave (PAW) method[49,50], as implemented in Vienna ab-initio simulation package (VASP)[51,52]. The Perdew-Burke-Ernzerhof (PBE) generalized gradient approximation exchange-correlations functional[53] was used. In the self-consistent process, $16 \times 4 \times 4$ k-point sampling grids were used, and the cut-off energy for plane wave expansion was 500 eV. The monolayer structure is extracted from the experimental bulk crystal, where inversion symmetry (IS) is slightly broken. With relaxation the symmetry of the relaxed monolayer becomes *Pmmn* (No. 59). The PBE band structure shows that there is a small band overlap at $\Gamma$. The band representation analysis of the monolayer band structure shows that there is a remarkable double-band inversion on the $\Gamma$- Y line. In order to match the experimental energy gap, we artificially upshift the conduction bands (Ta-$dz^2$ in the Wannier-based Hamiltonian) by 0.1 eV to remove the small band overlap.

**Growth of Ta$_2$Pd$_3$Te$_5$ crystals**

Single crystals of Ta$_2$Pd$_3$Te$_5$ were synthesized by self-flux method. Details can be found in Ref [29].

**Devices fabrication**

We have two kinds of fabrication for devices: one is coating devices with PMMA to prevent from surface oxidization and contamination, and using SiO$_2$/Si+ as bottom gate (devices #A1-A5, #A7-A14, and #B1-B6); the other is using a graphite/hexagonal boron nitride (hBN) bottom gate and covering devices with hBN on top to protect devices (device #A6, Supplementary device #1 and #2).

For the first kind of devices: The $Ta_2Pd_3Te_5$ thin films were mechanically exfoliated onto $SiO_2/Si^+$ substrates and coated with PMMA in a glove box. The two-terminal/four-terminal/Hall bar devices were patterned using electron-beam lithography and moved into the glove box to develop. The Ti/Au electrodes were deposited using thermal evaporation instrument installed in the glove box. Then lifted-off the metal film and coated with PMMA once again without being moved out of the glove box. After these, we performed electron-beam lithography for the second time to pattern the Ti/Au pads in order that we could connect the pads and transport measurements sample holder using wire-bonding out of the glove box. The $Ta_2Pd_3Te_5$ thin films were in the nitrogen atmosphere glove box or coated with PMMA in the air during the whole fabrication process to prevent oxidation.

For the second kind of devices: We used polydimethylsiloxane (PDMS) to exfoliate graphite/hBN/$Ta_2Pd_3Te_5$ flake and drop them on $SiO_2/Si^+$ substrates respectively on a micro-positioning stage in the glove box to build a graphite/hBN/$Ta_2Pd_3Te_5$ stack. Then coat the substrate with PMMA without moving out of the glove box. The two-terminal/four-terminal/Hall bar devices were patterned using electron-beam lithography and moved into the glove box to develop. The Ti/Au electrodes were deposited using thermal evaporation instrument installed in the glove box. Then lifted-off the metal film and coated the device with hBN once again without being moved out of the glove box.

**Electrical transport measurements**

The electrical transport measurements were performed in cryostats (Oxford instruments dilution refrigerator with temperature range ~10 mK - 30 K, Oxford instruments $^3$He cryostats with temperature range ~ 400 mK - 30K and Oxford instruments TeslatronPT with temperature range ~ 1.5 K - 300 K). The bottom gate voltage $V_{bg}$ and d.c. bias were applied using Keithley 2400 or 2612. Standard lock-in measurements were taken with a frequency of 3 - 30 Hz with a small a.c. excitation below the thermal excitation range without specially marked. Note that due to the disorders and defects which could break the conducting channels in the edges of $Ta_2Pd_3Te_5$, the tunneling behaviors appear both in the electrodes - edge junctions (FL - LL) and edge - edge junctions (LL - LL)[8, 36, 37]. So, we can observe the power law behavior of Luttinger liquid both in two-terminal and four-terminal/Hall bar configurations.

Note that due to the thickness of SiO$_2$ dielectric layers used to apply $V_{bg}$ are different in device #A1 (200 nm) and #A2 (100 nm), the adjusting abilities are different. The thickness of SiO$_2$ is 100 nm if not specially marked in other devices.

## Data availability

All data needed to evaluate the conclusions in the paper are present in the main text and/or the supplementary information. The authors declare that all of the raw data generated in this study have been deposited in Figshare (http://dx.doi.org/10.6084/m9.figshare.24407743).


## Acknowledgments

We are grateful to T. Xiang, J.P. Hu, Y. Zhou, Y. Wan, M. Cheng, B.J. Feng, and Z.Y. Sun for discussions. The work of J.S., L.L., F.Q. and G.L. were supported by the Beijing Natural Science Foundation (Grant No. JQ23022), the Strategic Priority Research Program B of Chinese Academy of Sciences (Grant No. XDB33000000), the Beijing Nova Program (Grant No. Z211100002121144), the National Natural Science Foundation of China (Grant Nos. 92065203 and 12174430), and the Synergetic Extreme Condition User Facility (SECUF). The work of other authors were supported by the National Natural Science Foundation of China (Grant Nos. 11974395, 12188101, U2032204, U1832202, U22A6005, 62022089, 11888101, 52072400 and 52025025), the Strategic Priority Research Program of the Chinese Academy of Sciences (Grant Nos. XDB33000000, XDB33030000 and XDB28000000), the National Key Research and Development Program of China (Grant Nos. 2022YFA1403800, 2021YFA1401800 and 2019YFA0308000), the Informatization Plan of Chinese Academy of Sciences (Grant No. CAS-WX2021SF-0102), the Chongqing Outstanding Youth Fund (Grant No. 2021ZX0400005), the China Postdoctoral Science Foundation (Grant Nos. 2021M703462, 2021TQ0356 and 2021M703461), the "Dreamline" beamline of Shanghai Synchrotron Radiation Facility (SSRF), the Center for Materials Genome, and the Synergetic Extreme Condition User Facility (SECUF).


## Author contributions

J. S. conceived and designed the experiment.

A.Q.W., Y.P.L., G.Y., Q.C.Z., D.G.Q., H.W., and X.C.G. fabricated devices and performed the transport measurements, supervised by G.T.L, F.M.Q., L.L., and J.S.

D.Y.Y. and Y.G.S. grew bulk $Ta_2Pd_3Te_5$ crystals.

F.Q.M., Q.H.Z., and L.G. performed the STEM measurements.

J.R.H. and T.Q. performed the ARPES measurements.

Z.P.G., J.C.G., and Z.J.W. performed the theoretical modeling.

H.Y. and X.J.Z. provided some supports on thin film exfoliation and devices fabrication.

A.Q.W., Y.P.L., G.Y., and J.S. analyzed the data and wrote the manuscript, with input from all authors.

**Competing interests**

Authors declare that they have no competing interests.

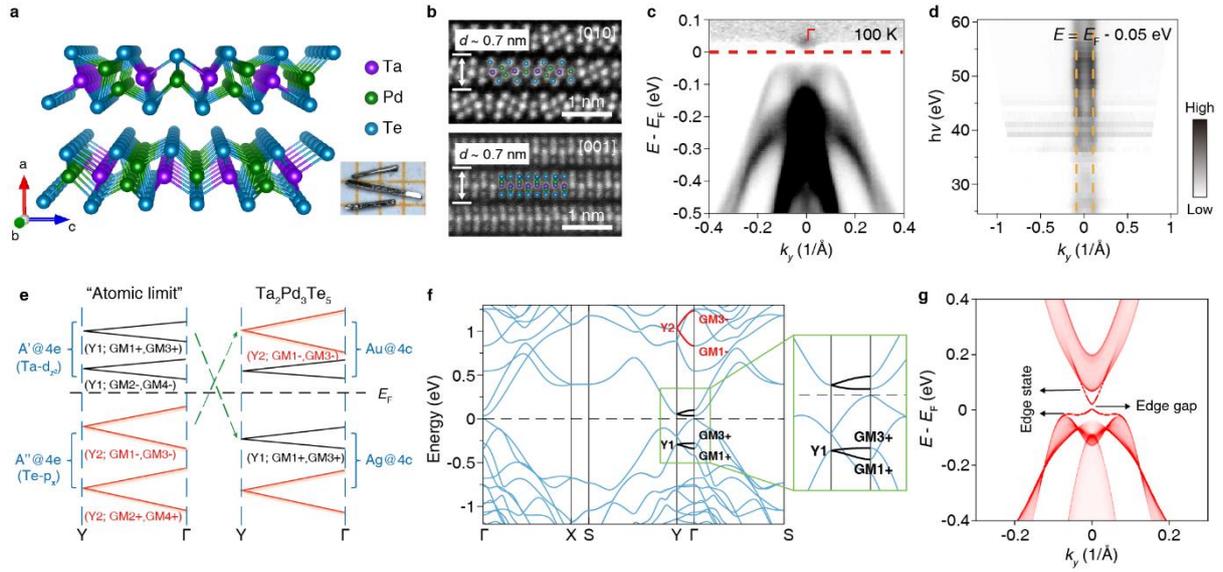

**Fig. 1 Crystal structure and band structure for $Ta_2Pd_3Te_5$. a,** Crystal structure viewed from the chain direction along the *b*-axis. Inset: Optical image of single crystals. Side length of the square is 1 mm. **b,** Cross-sectional STEM images of $Ta_2Pd_3Te_5$ single crystal along [010] (up) and [001] (down) direction. **c,** ARPES intensity plot of the band structure along the $\bar{\Gamma}$ - $\bar{Y}$ direction at $T = 100$ K. For clarity, the data is divided by the Fermi-Dirac distribution function to visualize the conduction band bottom above $E_F$. **d,** Intensity plot of ARPES data at $E = E_F -$ 0.05 eV (valence band maximum) collected in a range of photon energies from 25 to 60 eV which shows weak coupling between $Ta_2Pd_3Te_5$ vdW layers. **e,** The diagram of double-band inversion along Y – Γ line (adapted from Ref. 28). **f,** The modified band structure by adding 0.1 eV onsite energy on Ta-$d_{z^2}$ states to simulate the gapped band structure with interactions. **g,** The 1D gapped edge state of the 2D quadrupole topological insulator. The onsite energy is slightly modified in order to match the experimental energy gap.

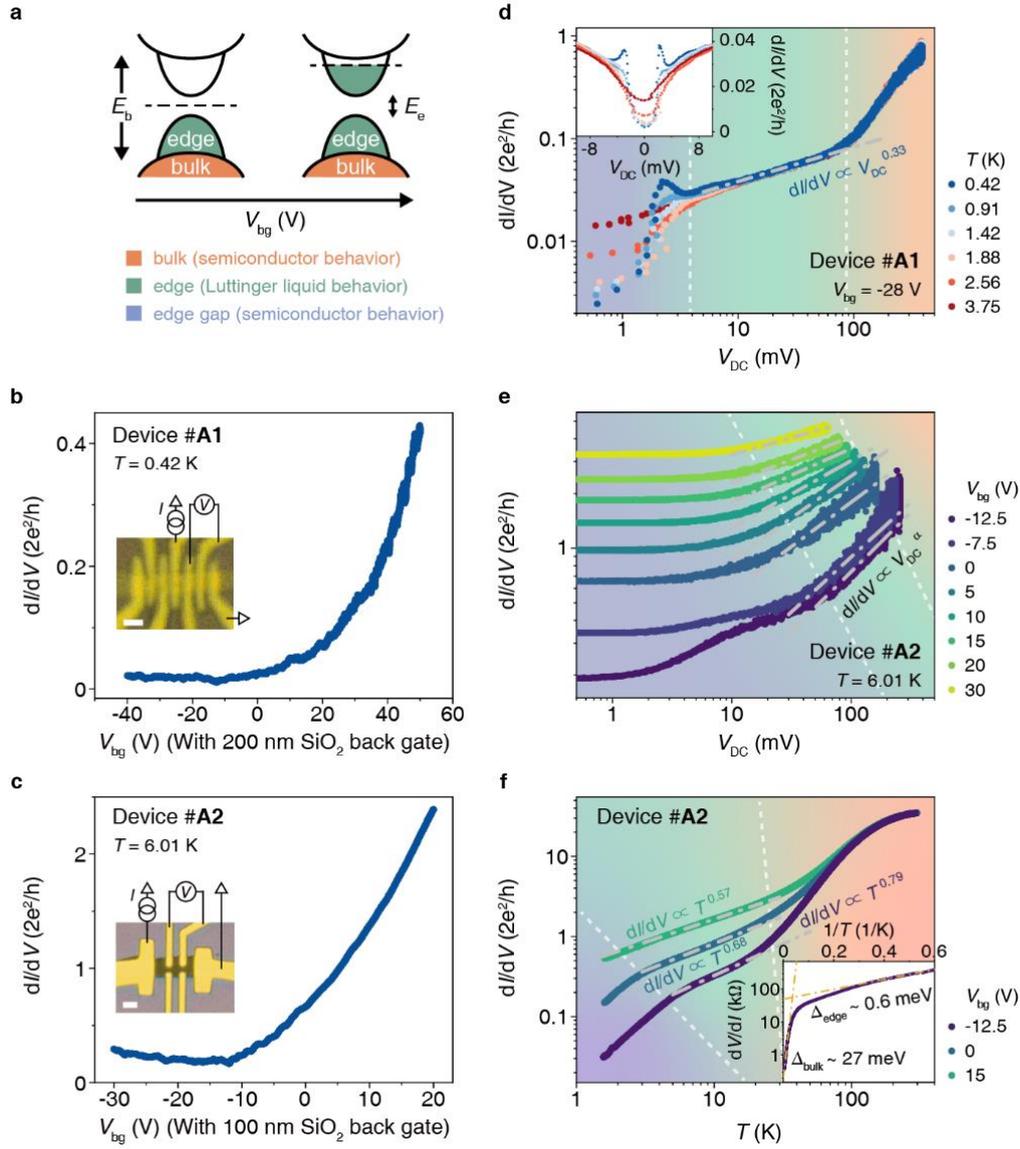

**Fig. 2 Electrical transport measurement for Ta$_2$Pd$_3$Te$_5$. a,** Sketch of Fermi level rising with $V_{bg}$. **b,** d$I$/d$V$ versus $V_{bg}$ measured in device #A1 at $T = 0.42$ K. Inset: Optical image and schematic measurement configuration. Scale bar, 1 μm. **c,** d$I$/d$V$ versus $V_{bg}$ measured in device #A2 at $T = 6.01$ K. Inset: Optical image and schematic measurement configuration. Scale bar, 2 μm. **d,** Log-log plot of d$I$/d$V$ versus $V_{DC}$ measured in device #A1 at $V_{bg} = -28$ V and different $T$. Inset: The same data as main figure but plotted in linear scale and symmetric $V_{DC}$. **e,** Log-log plot of d$I$/d$V$ versus $V_{DC}$ measured in device #A2 at $T = 6.01$ K and different $V_{bg}$. **f,** Log-log plot of temperature dependence d$I$/d$V$ of device #A2 at different $V_{bg}$. Inset: log (d$V$/d$I$) v.s. 1/$T$ plot of the temperature dependence resistance of device #A2 at $V_{bg} = -12.5$ V. The gray dot

dash lines show the power-law behavior. The orange/green/purple region of the background shows the bulk semiconductor/edge Luttinger liquid/edge gap behavior.

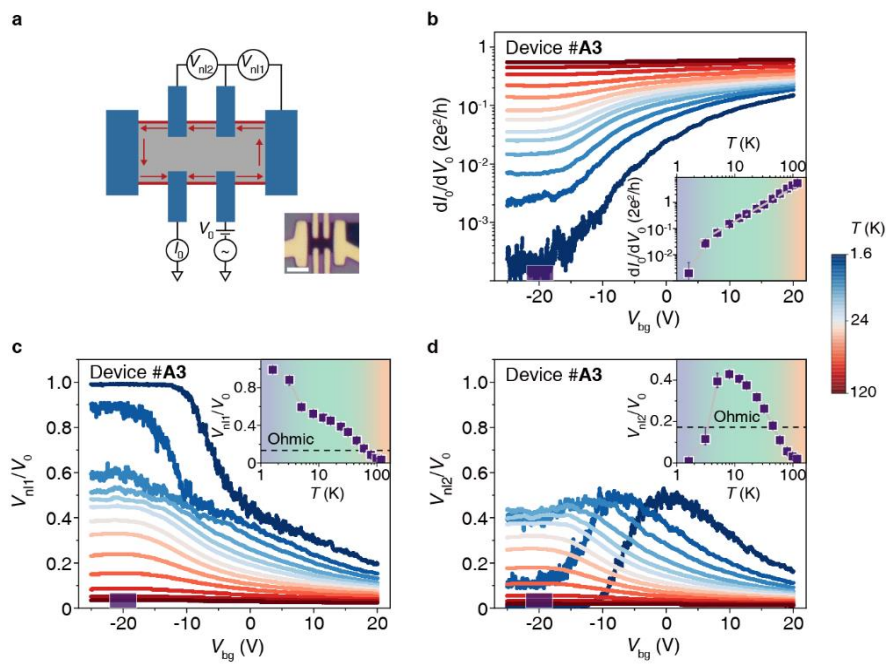

**Fig. 3 Nonlocal transport and edge states in Ta$_2$Pd$_3$Te$_5$. a,** Schematic nonlocal measurement configuration with Hall bar shape contacts (blue) on a Ta$_2$Pd$_3$Te$_5$ thin film (gray). The red lines are edges of the thin film and red arrows are current flow along edges. Inset: Optical image of device #A3. Scale bar, 2 μm. **b,** Local differential conductance d$I_0$/d$V_0$ measured in device #A3 as a function of $V_{bg}$ at differential $T$. **c,d,** Nonlocal voltage ratio $V_{nl1}/V_0$ **(c)** and $V_{nl2}/V_0$ **(d)** versus $V_{bg}$ at different $T$. Inset in **b**, **c** and **d**: d$I_0$/d$V_0$ **(b)**, $V_{nl1}/V_0$ **(c)** and $V_{nl2}/V_0$ **(d)** versus $T$ taken as the mean value at $V_{bg}$ = -22 V to -18 V. The error bars show the full range of values spanned by the fluctuations. Dotted-lines are labels with simulation values of homogeneous 2D ohmic resistivity situation.

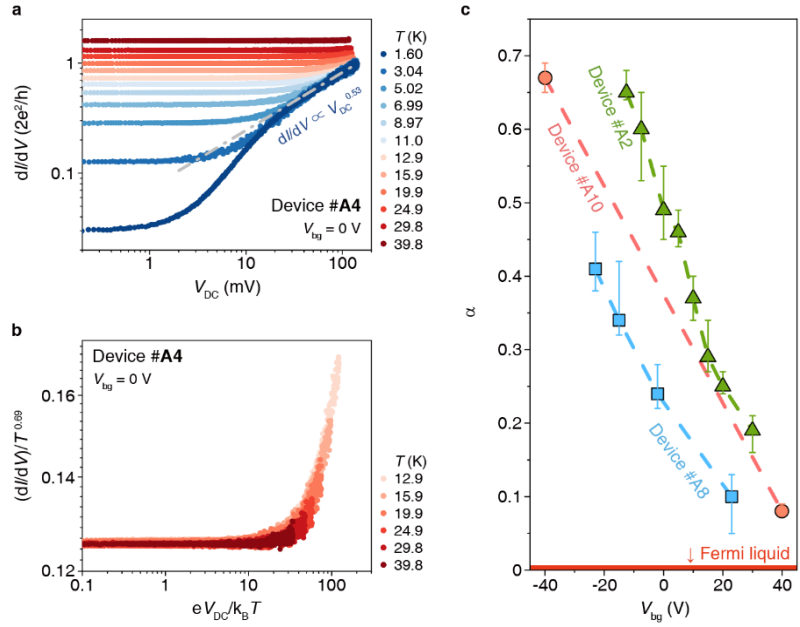

**Fig. 4 Tunable Luttinger liquid behaviors. a,** d$I$/d$V$ versus $V_{DC}$ measured in device #A4 at differential $T$ with $V_{bg}$ = 0 V. **b,** The d$I$/d$V$($V_{DC}$) curves without edge gap behavior in **(a)** (From $T$ = 12.9 K to 39.8 K) are plotted as scaled conductance (d$I$/d$V$)/$T^{0.69}$ versus scaled temperature e$V_{DC}$/k$_B T$. All data collapses to a single curve indicating Luttinger liquid behaviors. (d$I$/d$V$($V_{DC}$) curves below $T$ = 12.9 K exhibit edge gap behavior due to the 2D SOTI nature, so they shouldn't obey the scaling behavior physically.) **c,** Gate tunable α in Luttinger liquid behaviors in d$I$/d$V$ v.s. $V_{DC}$ curves. The error bars reflect the uncertainty due to measurement noise. (Raw data are included in Supplementary Fig. 12.)

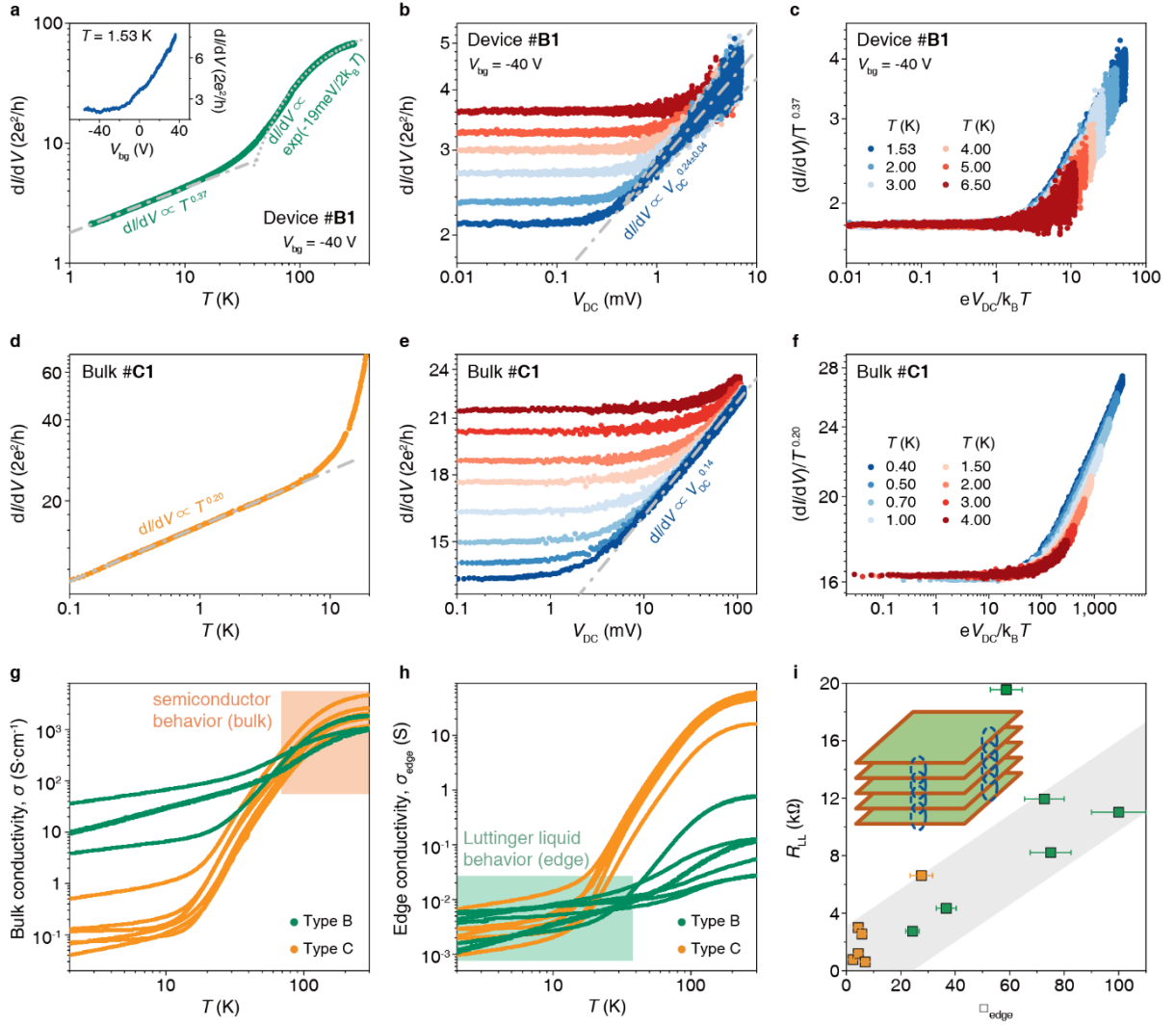

**Fig. 5 Universal Luttinger liquid behaviors from atomic to macroscopic scale. a,** Temperature dependence $dI/dV$ measured in device #B1 at $V_{bg}$ = -40 V. The gray dot dash line and gray dot line show the power-law behavior and semiconductor behavior respectively. Inset: $dI/dV$ versus $V_{bg}$ at $T$ = 1.53 K. **b,** $dI/dV$ versus $V_{DC}$ measured in device #B1 at $V_{bg}$ = -40 V and different $T$. Due to the noise in the large bias region of $dI/dV(V_{DC})$ curves, we use the median values to extract the exponent of power-law behavior (dot dash line) and use the upper and lower envelopes as fitting errors (dash lines)[10]. **c,** The same data as **(b)**, but plotted as scaled conductance $(dI/dV)/T^{0.37}$ versus scaled temperature $eV_{DC}/k_BT$. All data collapses to a single curve. **d,** $dI/dV$ versus $T$ measured in bulk sample #C1. **e,** $dI/dV$ versus $V_{DC}$ measured in bulk sample #C1 at different $T$. **f,** The same data as **(e)**, but plotted as scaled conductance $(dI/dV)/T^{0.20}$ versus scaled temperature $eV_{DC}/k_BT$. **g,h,** Temperature dependence of conductivity $\sigma$ **(g)** and edge conductivity $\sigma_{edge}$ (**h**, defined as conductance / thickness × length) for type B (green line) and C (orange line) samples. Three devices measured with two-terminal configuration are not included in **(g)**, others are four-terminal configuration. **i,** Resistances of starting points of Luttinger liquid behaviors $R_{LL}$ versus side surface square number $\square_{edge}$ (defined as length / thickness) for type B and C samples, which follows an approximate linear dependence. The error bars reflect the uncertainty from length / thickness measurement. Uncertainties in y-axis are smaller than point size. Inset: Schematic of vdW stacking layers in $Ta_2Pd_3Te_5$. The red lines are edges of each layer and blue dotted lines are sketches of weak coupling between layers.

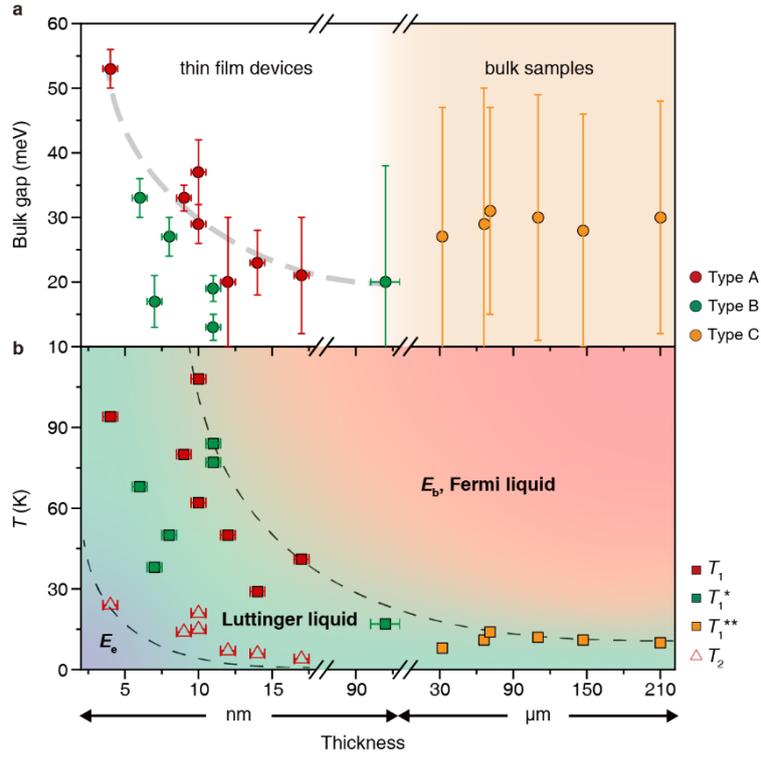

**Fig. 6 Luttinger liquid phase diagrams. a,** Thickness-dependent bulk gap extracted from d$I$/d$V$($T$) curves. The error bars in $x$-axis reflect the uncertainty from thickness measurement and error bars in $y$-axis show the ranges of bulk gap variation. **b,** The Luttinger liquid phase transition temperature versus sample thickness. $T_1$/$T_1^*$/$T_1^{**}$: the transition temperatures from bulk Fermi-liquid semiconductor behaviors to edge Luttinger liquid behaviors for type A/B/C samples; $T_2$: the transition temperatures from edge Luttinger liquid behaviors to edge gap behaviors for type A samples. The error bars reflect the uncertainty from thickness measurement. Uncertainties in $y$-axis are smaller than point size.


**References**

1. Tokura, Y. Quantum materials at the crossroads of strong correlation and topology. *Nat. Mater.* **21**, 971–973 (2022).

2. Zhou, Y., Kanoda, K. & Ng, T.-K. Quantum spin liquid states. *Rev. Mod. Phys.* **89**, 025003 (2017).

3. Raghu, S., Qi, X.-L., Honerkamp, C. & Zhang, S.-C. Topological Mott Insulators. *Phys. Rev. Lett.* **100**, 156401 (2008).

4. Cao, Y. *et al.* Correlated insulator behaviour at half-filling in magic-angle graphene superlattices. *Nature* **556**, 80–84 (2018).

5. Xie, M. & MacDonald, A. H. Nature of the Correlated Insulator States in Twisted Bilayer Graphene. *Phys. Rev. Lett.* **124**, 097601 (2020).

6. Hasan, M. Z. & Kane, C. L. Colloquium: Topological insulators. *Rev. Mod. Phys.* **82**, 23 (2010).

7. Qi, X.-L. & Zhang, S.-C. Topological insulators and superconductors. *Rev. Mod. Phys.* **83**, 54 (2011).

8. Li, T. *et al.* Observation of a Helical Luttinger Liquid in InAs/GaSb Quantum Spin Hall Edges. *Phys. Rev. Lett.* **115**, 136804 (2015).

9. Strunz, J. *et al.* Interacting topological edge channels. *Nat. Phys.* **16**, 83–88 (2020).

10. Stühler, R. *et al.* Tomonaga–Luttinger liquid in the edge channels of a quantum spin Hall insulator. *Nat. Phys.* **16**, 47–51 (2020).

11. Milliken, F. P., Umbach, C. P. & Webb, R. A. Indications of a Luttinger liquid in the fractional quantum Hall regime. *Solid State Commun.* **97**, 309–313 (1996).

12. Chang, A. M., Pfeiffer, L. N. & West, K. W. Observation of Chiral Luttinger Behavior in Electron Tunneling into Fractional Quantum Hall Edges. *Phys. Rev. Lett.* **77**, 2538–2541 (1996).

13. Hilke, M., Tsui, D. C., Grayson, M., Pfeiffer, L. N. & West, K. W. Fermi Liquid to Luttinger Liquid Transition at the Edge of a Two-Dimensional Electron Gas. *Phys. Rev. Lett.* **87**, 186806 (2001).

14. Chang, A. M. Chiral Luttinger liquids at the fractional quantum Hall edge. *Rev. Mod. Phys.* **75**, 1449–1505 (2003).

15. Hashisaka, M., Hiyama, N., Akiho, T., Muraki, K. & Fujisawa, T. Waveform measurement


of charge- and spin-density wavepackets in a chiral Tomonaga–Luttinger liquid. *Nat. Phys.* **13**, 559–562 (2017).

16. Randeria, M. T. *et al.* Interacting multi-channel topological boundary modes in a quantum Hall valley system. *Nature* **566**, 363–367 (2019).

17. Wen, X. G. Chiral Luttinger liquid and the edge excitations in the fractional quantum Hall states. *Phys. Rev. B* **41**, 12838–12844 (1990).

18. Kane, C. L. & Fisher, M. P. A. Transport in a one-channel Luttinger liquid. *Phys. Rev. Lett.* **68**, 1220–1223 (1992).

19. Maciejko, J. *et al.* Kondo Effect in the Helical Edge Liquid of the Quantum Spin Hall State. *Phys. Rev. Lett.* **102**, 256803 (2009).

20. Imambekov, A., Schmidt, T. L. & Glazman, L. I. One-dimensional quantum liquids: Beyond the Luttinger liquid paradigm. *Rev. Mod. Phys.* **84**, 1253–1306 (2012).

21. Giamarchi, T. Some experimental tests of Tomonaga-Luttinger liquids. *Int. J. Mod. Phys. B* **26**, 1244004 (2012).

22. Xie, B. *et al.* Higher-order band topology. *Nat. Rev. Phys.* **3**, 520–532 (2021).

23. Ezawa, M. Higher-Order Topological Insulators and Semimetals on the Breathing Kagome and Pyrochlore Lattices. *Phys. Rev. Lett.* **120**, 026801 (2018).

24. Kunst, F. K., van Miert, G. & Bergholtz, E. J. Lattice models with exactly solvable topological hinge and corner states. *Phys. Rev. B* **97**, 241405 (2018).

25. Park, M. J., Kim, Y., Cho, G. Y. & Lee, S. Higher-Order Topological Insulator in Twisted Bilayer Graphene. *Phys. Rev. Lett.* **123**, 216803 (2019).

26. Sheng, X.-L. *et al.* Two-Dimensional Second-Order Topological Insulator in Graphdiyne. *Phys. Rev. Lett.* **123**, 256402 (2019).

27. Ren, Y., Qiao, Z. & Niu, Q. Engineering Corner States from Two-Dimensional Topological Insulators. *Phys. Rev. Lett.* **124**, 166804 (2020).

28. Guo, Z., Deng, J., Xie, Y. & Wang, Z. Quadrupole topological insulators in $Ta_2M_3Te_5$ ($M$=Ni, Pd) monolayers. Npj Quantum Mater. 7, 1–6 (2022).

29. Guo, Z. *et al.* Quantum spin Hall effect in $Ta_2M_3Te_5$ ($M$=Pd, Ni). *Phys. Rev. B* **103**, 115145 (2021).

30. Wang, X. *et al.* Observation of topological edge states in the quantum spin Hall insulator

Ta$_2$Pd$_3$Te$_5$. *Phys. Rev. B* **104**, L241408 (2021).

31. Gao, J. *et al.* Unconventional materials: the mismatch between electronic charge centers and atomic positions. *Sci. Bull.* **67**, 598–608 (2022).

32. Xu, Y. *et al.* Filling-Enforced Obstructed Atomic Insulators. *ArXiv:2106.10276* (2021).

33. Fei, Z. *et al.* Edge conduction in monolayer WTe$_2$. *Nat. Phys.* **13**, 677–682 (2017).

34. Aharon-Steinberg, A. *et al.* Long-range nontopological edge currents in charge-neutral graphene. *Nature* **593**, 528–534 (2021).

35. Bockrath, M. *et al.* Luttinger-liquid behaviour in carbon nanotubes. *Nature* **397**, 598–601 (1999).

36. Yao, Z., Postma, H. W. C., Balents, L. & Dekker, C. Carbon nanotube intramolecular junctions. *Nature* **402**, 273–276 (1999).

37. Hsu, C.-H. *et al.* Charge transport of a spin-orbit-coupled Luttinger liquid. *Phys. Rev. B* **100**, 195423 (2019).

38. Grayson, M., Tsui, D. C., Pfeiffer, L. N., West, K. W. & Chang, A. M. Continuum of Chiral Luttinger Liquids at the Fractional Quantum Hall Edge. *Phys. Rev. Lett.* **80**, 1062–1065 (1998).

39. Venkataraman, L., Hong, Y. S. & Kim, P. Electron Transport in a Multichannel One-Dimensional Conductor: Molybdenum Selenide Nanowires. *Phys. Rev. Lett.* **96**, 076601 (2006).

40. Wang, J. *et al.* A tied Fermi liquid to Luttinger liquid model for nonlinear transport in conducting polymers. *Nat. Commun.* **12**, 58 (2021).

41. Blumenstein, C. *et al.* Atomically controlled quantum chains hosting a Tomonaga–Luttinger liquid. *Nat. Phys.* **7**, 776–780 (2011).

42. Yi, W., Lu, L., Hu, H., Pan, Z. W. & Xie, S. S. Tunneling into Multiwalled Carbon Nanotubes: Coulomb Blockade and the Fano Resonance. *Phys. Rev. Lett.* **91**, 076801 (2003).

43. Fogler, M. M., Malinin, S. V. & Nattermann, T. Coulomb Blockade and Transport in a Chain of One-Dimensional Quantum Dots. *Phys. Rev. Lett.* **97**, 096601 (2006).

44. Rodin, A. S. & Fogler, M. M. Apparent Power-Law Behavior of Conductance in Disordered Quasi-One-Dimensional Systems. *Phys. Rev. Lett.* **105**, 106801 (2010).

45. Li, Y. P. *et al*. Interfering Josephson diode effect and magnetochiral anisotropy in Ta$_2$Pd$_3$Te$_5$ asymmetric edge interferometer. *Arxiv:* 2306.08478 (2023).

46. Zaliznyak, I. A. A glimpse of a Luttinger liquid. *Nat. Mater.* **4**, 273–275 (2005).


47. Lake, B., Tennant, D. A., Frost, C. D. & Nagler, S. E. Quantum criticality and universal scaling of a quantum antiferromagnet. *Nat. Mater.* **4**, 329–334 (2005).

48. Higashihara, N. *et al.* Superconductivity in $Nb_2Pd_3Te_5$ and Chemically-Doped $Ta_2Pd_3Te_5$. *J. Phys. Soc. Jpn.* **90**, 063705 (2021).

49. Blöchl, P. E. Projector augmented-wave method. *Phys. Rev. B* **50**, 17953–17979 (1994).

50. Kresse, G. & Joubert, D. From ultrasoft pseudopotentials to the projector augmented-wave method. *Phys. Rev. B* **59**, 1758–1775 (1999).

51. Kresse, G. & Furthmüller, J. Efficiency of ab-initio total energy calculations for metals and semiconductors using a plane-wave basis set. *Comput. Mater. Sci.* **6**, 15–50 (1996).

52. Kresse, G. & Furthmüller, J. Efficient iterative schemes for *ab initio* total-energy calculations using a plane-wave basis set. *Phys. Rev. B* **54**, 11169–11186 (1996).

53. Perdew, J. P., Burke, K. & Ernzerhof, M. Generalized Gradient Approximation Made Simple. *Phys. Rev. Lett.* **77**, 3865–3868 (1996).


# Supplementary Information for

# A robust and tunable Luttinger liquid in correlated edge of transition-metal second-order topological insulator Ta$_2$Pd$_3$Te$_5$


Anqi Wang[1,2,†], Yupeng Li[1,†], Guang Yang[1,†], Dayu Yan[1,†], Yuan Huang[3], Zhaopeng Guo[1], Jiacheng Gao[1,2], Jierui Huang[1,2], Qiaochu Zeng[1], Degui Qian[1], Hao Wang[1], Xingchen Guo[1,2], Fanqi Meng[1], Qinghua Zhang[1,4], Lin Gu[1,2,5], Xingjiang Zhou[1,2,5], Guangtong Liu[1,5], Fanming Qu[1,2,5], Tian Qian[1,5], Youguo Shi[1,2,5*], Zhijun Wang[1,2*], Li Lu[1,2,5*], Jie Shen[1,5*]

[1]Beijing National Laboratory for Condensed Matter Physics, Institute of Physics, Chinese Academy of Sciences, Beijing 100190, China

[2]School of Physical Sciences, University of Chinese Academy of Sciences, Beijing 100049, China

[3]Advanced Research Institute of Multidisciplinary Science, Beijing Institute of Technology, Beijing 100081, China

[4]Yangtze River Delta Physics Research Center Co. Ltd, Liyang 213300, China

[5]Songshan Lake Materials Laboratory, Dongguan 523808, China

†These authors contributed equally to this work

*Corresponding author. Email: ygshi@iphy.ac.cn (Y.-G.S.), wzj@iphy.ac.cn (Z.-J.W.), lilu@iphy.ac.cn (L.L.), shenjie@iphy.ac.cn (J.S.)


**Supplementary Table 1. The irreps table of Gamma point of space group 59.**

|  | GM1+ | GM1- | GM2+ | GM2- | GM3+ | GM3- | GM4+ | GM4- |
|---|---|---|---|---|---|---|---|---|
| $\{E\|0,0,0\}$ | 1 | 1 | 1 | 1 | 1 | 1 | 1 | 1 |
| $\{C_{2x}\|0,1/2,1/2\}$ | 1 | 1 | 1 | 1 | -1 | -1 | -1 | -1 |
| $\{C_{2y}\|0,1/2,0\}$ | 1 | 1 | -1 | -1 | 1 | 1 | -1 | -1 |
| $\{C_{2z}\|0,0,1/2\}$ | 1 | 1 | -1 | -1 | -1 | -1 | 1 | 1 |
| $\{-1\|0,0,0\}$ | 1 | -1 | 1 | -1 | 1 | -1 | 1 | -1 |
| $\{M_x\|0,1/2,1/2\}$ | 1 | -1 | 1 | -1 | -1 | 1 | -1 | 1 |
| $\{M_y\|0,1/2,0\}$ | 1 | -1 | -1 | 1 | 1 | -1 | -1 | 1 |
| $\{M_z\|0,0,1/2\}$ | 1 | -1 | -1 | 1 | -1 | 1 | 1 | -1 |

**Supplementary Table 2. The irreps table of Y point of space group 59.**

| | Y1 | Y2 |
|---|---|---|
| $\{E\|0,0,0\}$ | $\begin{pmatrix}1 & 0\\0 & 1\end{pmatrix}$ | $\begin{pmatrix}1 & 0\\0 & 1\end{pmatrix}$ |
| $\{C_{2x}\|0,1/2,1/2\}$ | $\begin{pmatrix}0 & 1\\1 & 0\end{pmatrix}$ | $\begin{pmatrix}0 & 1\\1 & 0\end{pmatrix}$ |
| $\{C_{2y}\|0,1/2,0\}$ | $\begin{pmatrix}1 & 0\\0 & -1\end{pmatrix}$ | $\begin{pmatrix}-1 & 0\\0 & 1\end{pmatrix}$ |
| $\{C_{2z}\|0,0,1/2\}$ | $\begin{pmatrix}0 & -1\\1 & 0\end{pmatrix}$ | $\begin{pmatrix}0 & 1\\-1 & 0\end{pmatrix}$ |
| $\{-1\|0,0,0\}$ | $\begin{pmatrix}1 & 0\\0 & -1\end{pmatrix}$ | $\begin{pmatrix}1 & 0\\0 & -1\end{pmatrix}$ |
| $\{M_x\|0,1/2,1/2\}$ | $\begin{pmatrix}0 & -1\\1 & 0\end{pmatrix}$ | $\begin{pmatrix}0 & -1\\1 & 0\end{pmatrix}$ |
| $\{M_y\|0,1/2,0\}$ | $\begin{pmatrix}1 & 0\\0 & 1\end{pmatrix}$ | $\begin{pmatrix}-1 & 0\\0 & -1\end{pmatrix}$ |
| $\{M_z\|0,0,1/2\}$ | $\begin{pmatrix}0 & 1\\1 & 0\end{pmatrix}$ | $\begin{pmatrix}0 & -1\\-1 & 0\end{pmatrix}$ |

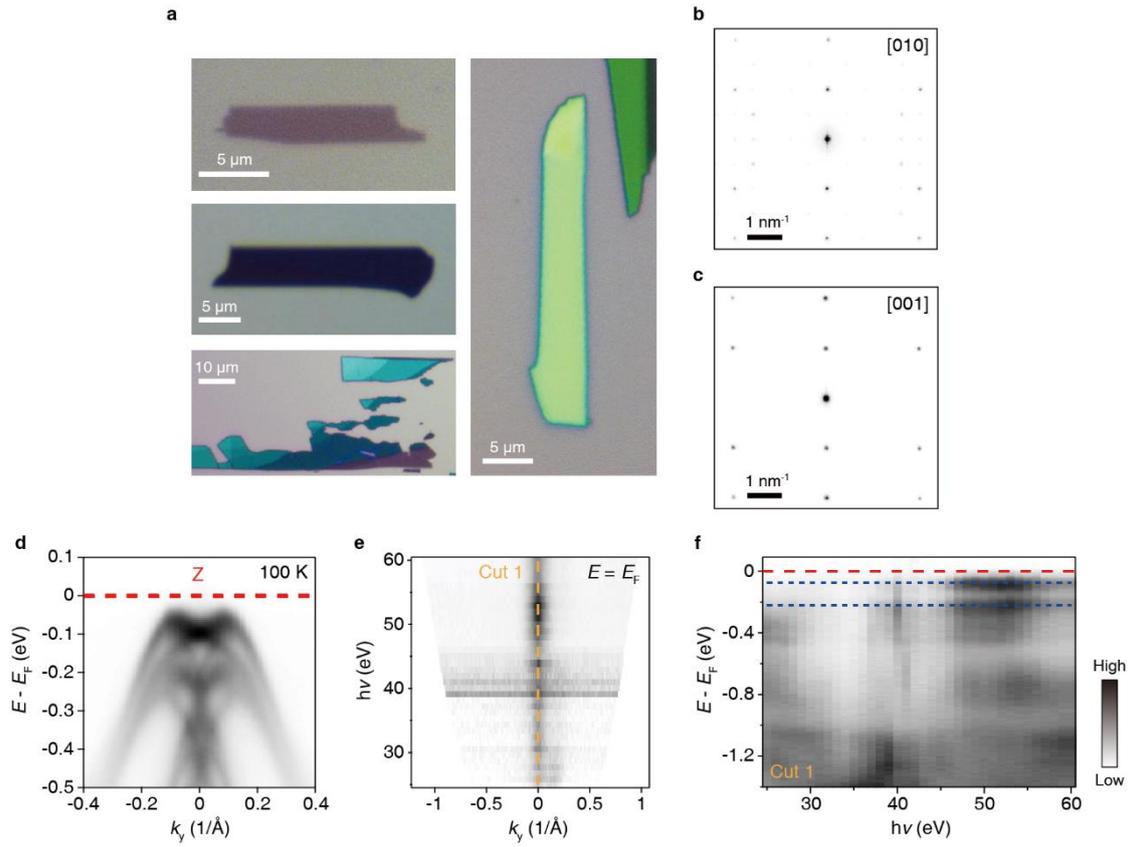

**Supplementary Fig. 1 Optical image, STEM image and ARPES data for $Ta_2Pd_3Te_5$. a,** Optical images of $Ta_2Pd_3Te_5$ thin films with different thicknesses. The uniform edges indicate the quasi-one-dimensional nature of $Ta_2Pd_3Te_5$. **b,c,** Fast Fourier transformed STEM images of $Ta_2Pd_3Te_5$ single crystal along [010] **(b)** and [001] **(c)** direction which give lattice parameters a ~ 13.9 Å and b ~ 3.7 Å. **d,** ARPES intensity plot of the band structure along the $\bar{Z}$ - $\bar{T}$ direction at $T$ = 100 K. **e,** Intensity plot of ARPES data at $E = E_F$ collected in a range of photon energies from 25 to 60 eV. **f,** Intensity plot of ARPES data along cut 1.

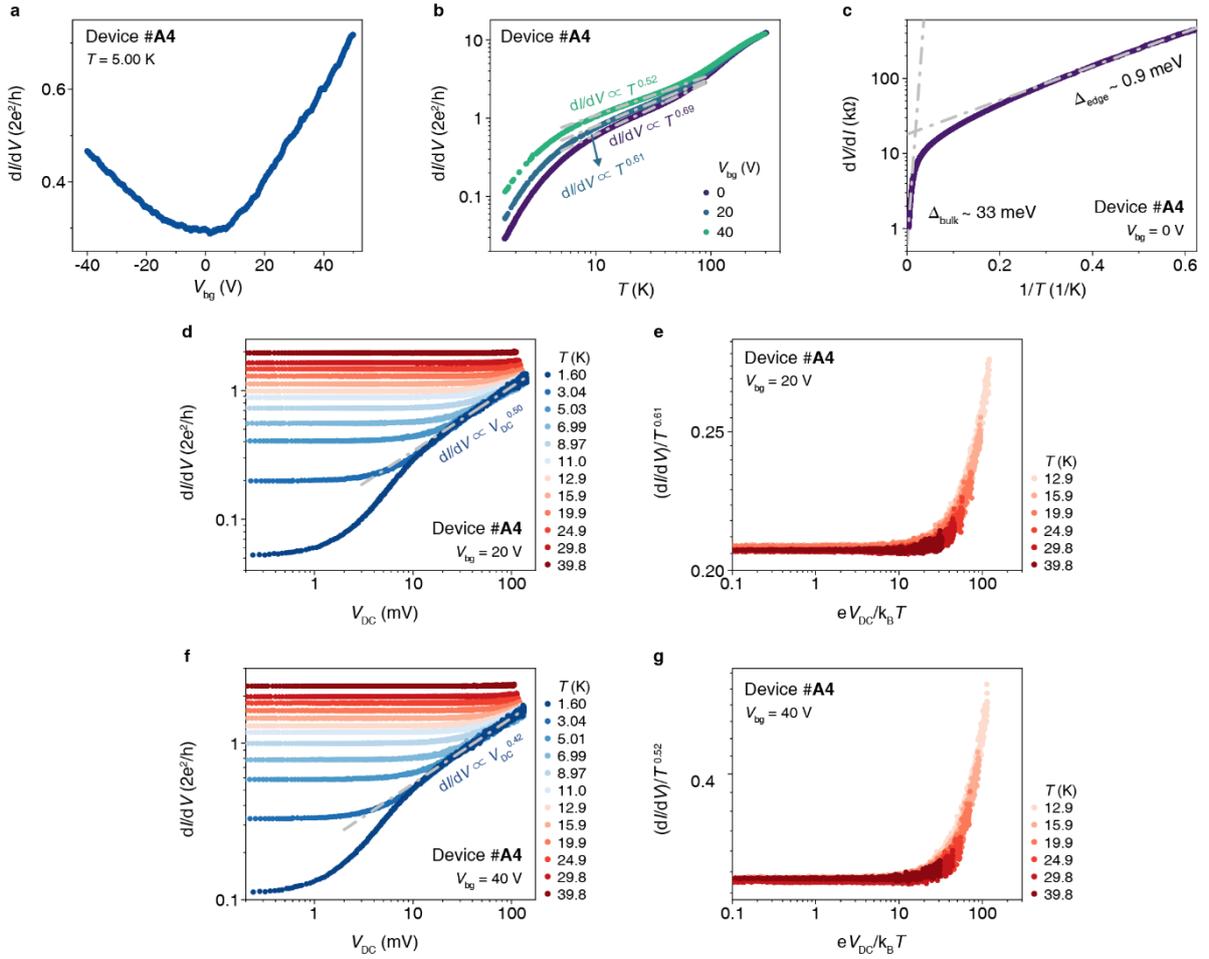

**Supplementary Fig. 2 Electrical transport measurement for device #A4. a,** d$I$/d$V$ versus $V_{bg}$ measured at $T = 5.00$ K. **b,** Log-log plot of temperature dependence d$I$/d$V$ at different $V_{bg}$. The gray dot dash lines show the power-law behaviors. **c,** log (d$V$/d$I$) v.s. $1/T$ plot of the temperature dependence resistance at $V_{bg} = 0$ V. **d,** d$I$/d$V$ versus $V_{DC}$ measured at differential $T$ with $V_{bg} = 20$ V. **e,** Curves form $T = 12.9$ K to 39.8 K in **(d)** are plotted as scaled conductance (d$I$/d$V$)/$T^{0.61}$ versus scaled temperature $eV_{DC}/k_BT$. **f,** d$I$/d$V$ versus $V_{DC}$ measured at differential $T$ with $V_{bg} = 40$ V. **g,** Curves form $T = 12.9$ K to 39.8 K in **(f)** are plotted as scaled conductance (d$I$/d$V$)/$T^{0.52}$ versus scaled temperature $eV_{DC}/k_BT$.

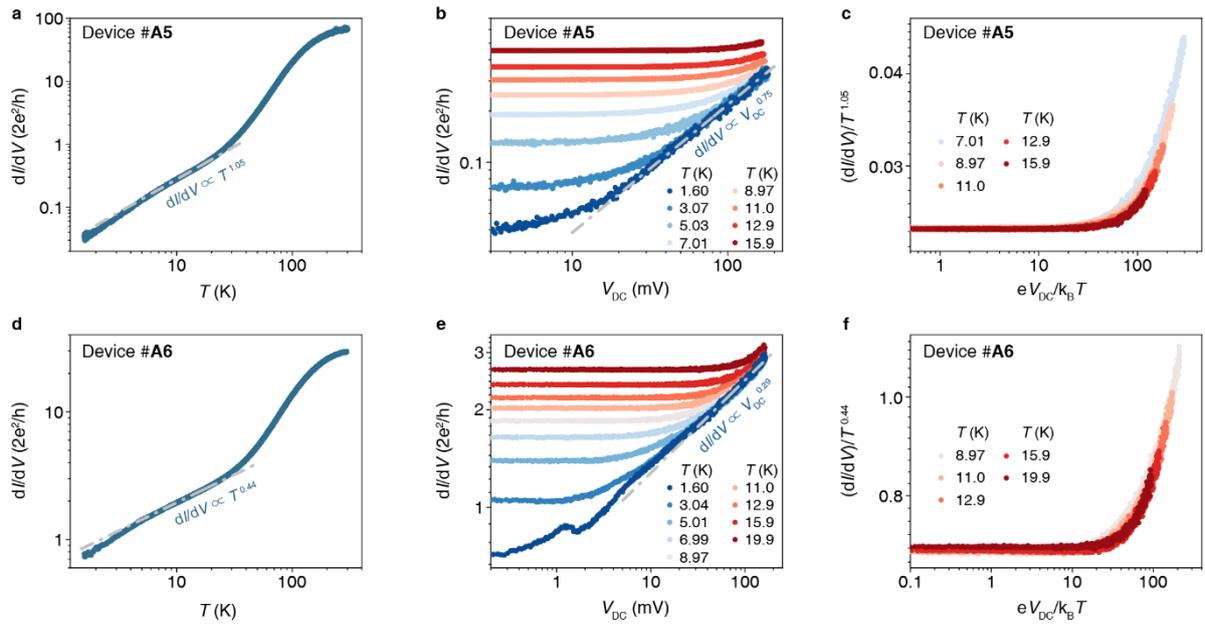

**Supplementary Fig. 3 Luttinger liquid behaviors in devices #A5 and #A6. a,** Temperature dependence d$I$/d$V$ of device #A5. **b,** d$I$/d$V$ versus $V_{DC}$ of device #A5 at differential $T$. **c,** Curves form $T$ = 7.01 K to 15.9 K in (**b**) are plotted as scaled conductance (d$I$/d$V$)/$T^{1.05}$ versus scaled temperature e$V_{DC}$/k$_B$$T$. **d,** Temperature dependence d$I$/d$V$ of device #A6. **e,** d$I$/d$V$ versus $V_{DC}$ of device #A6 at differential $T$. **f,** Curves form $T$ = 8.97 K to 19.9 K in (**e**) are plotted as scaled conductance (d$I$/d$V$)/$T^{0.44}$ versus scaled temperature e$V_{DC}$/k$_B$$T$.

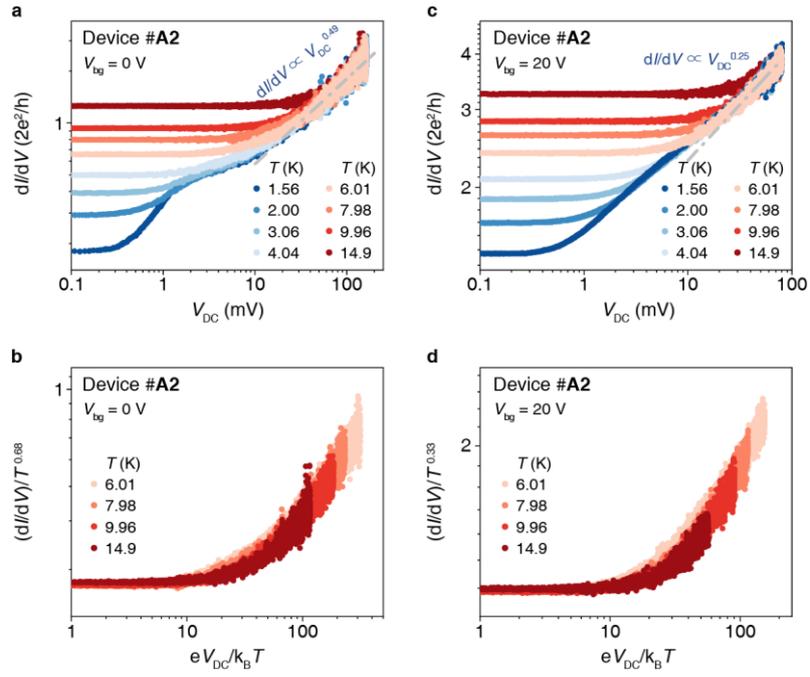

**Supplementary Fig. 4 Luttinger liquid behavior in device #A2. a,** d$I$/d$V$ versus $V_{DC}$ at differential $T$ with $V_{bg}$ = 0 V. **b,** Curves form $T$ = 6.01 K to 14.9 K in (**a**) are plotted as scaled conductance (d$I$/d$V$)/$T^{0.68}$ versus scaled temperature e$V_{DC}$/k$_B T$. **c,** d$I$/d$V$ versus $V_{DC}$ at differential $T$ with $V_{bg}$ = 20 V. **d,** Curves form $T$ = 6.01 K to 14.9 K in (**c**) are plotted as scaled conductance (d$I$/d$V$)/$T^{0.33}$ versus scaled temperature e$V_{DC}$/k$_B T$.

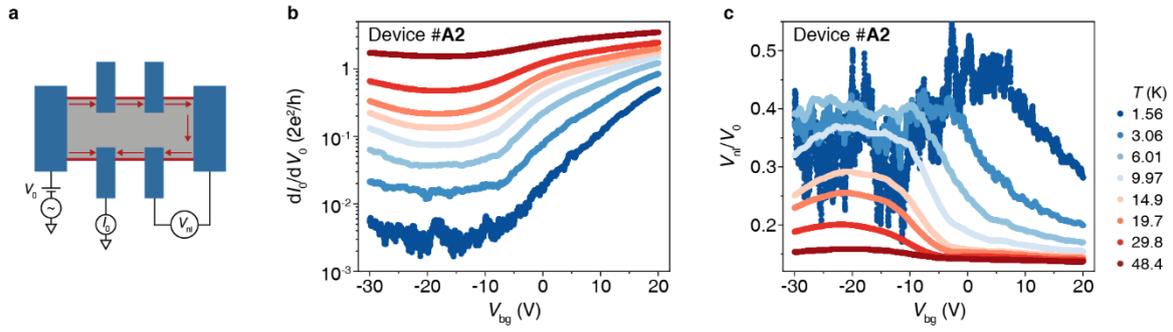

**Supplementary Fig. 5 Nonlocal measurement for device #A2. a,** Schematic nonlocal measurement configuration with Hall bar shape contacts (blue) on a $Ta_2Pd_3Te_5$ thin film (gray). The red lines are edges of the thin film and red arrows are current flow along edges. **b,** Local differential conductance $dI_0/dV_0$ measured in device #A2 as a function of $V_{bg}$ at differential $T$. **c,** Nonlocal voltage ratio $V_{nl}/V_0$ versus $V_{bg}$ at different $T$.

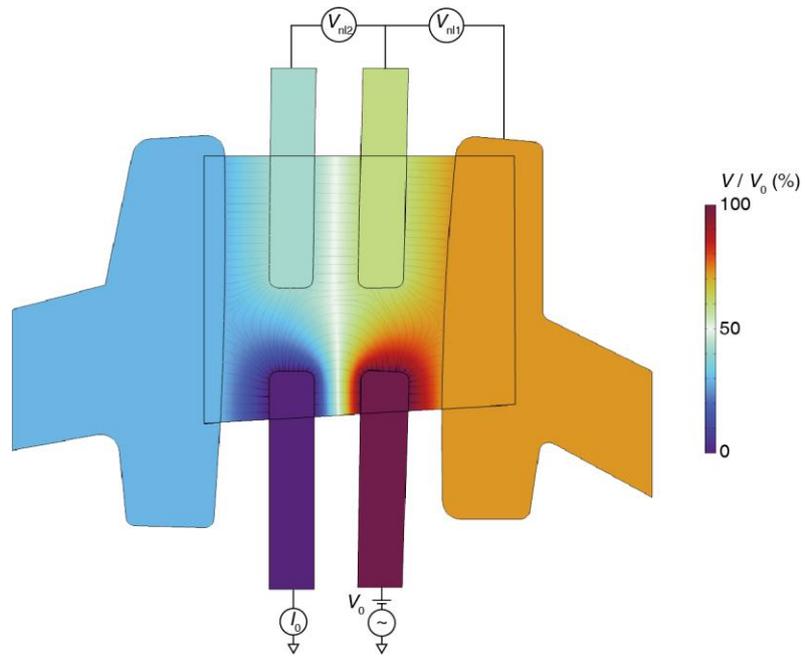

**Supplementary Fig. 6 Simulation of potential distribution in homogeneous 2D ohmic resistivity situation for device #A3**. It gives $V_{nl1}/V_0 \sim 13.2\%$ and $V_{nl2}/V_0 \sim 17.3\%$.

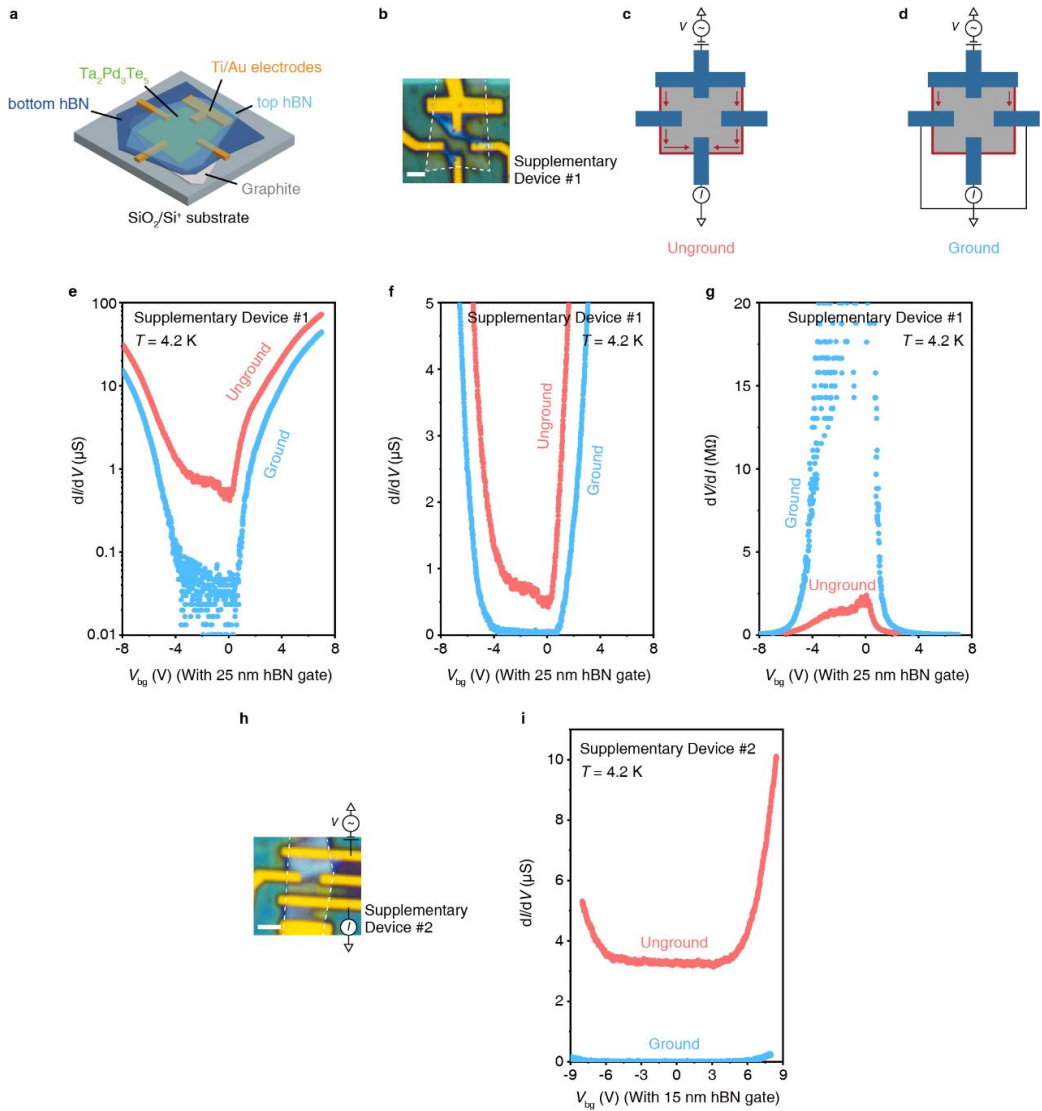

**Supplementary Fig. 7 Validation of edge state. a,** Sketch map of the hBN bottom gate devices (see 'Devices fabrication' section in Methods for detail). **b,** Optical image of Supplementary device #1 with hBN bottom gate. The dashed white line highlights the $Ta_2Pd_3Te_5$ thin film. The contrast color regions are due to the unevenness of the top hBN induced by the thick (~ 70 nm) Ti/Au electrodes which are insulating and serve only as protecting layers. So, they do not influence the transport behavior. (All other devices except device #A6 are coated with PMMA as protecting layer with clean and smooth surfaces indicating the high quality, e.g. inset in Fig. 2b,c, Fig. 3a, etc.) Scale bar, 2 μm. **c,d,** Schematic measurement configuration of the unground **(c)** and ground **(d)** configuration. The red lines are edges of the thin film and red arrows are current flow along edges. The edges are short out when grounding the side contacts and current can only flow through the bulk. **e,** $dI/dV$ versus $V_{bg}$ in unground (red) and ground (blue) configuration, the conductance around CNP is finite in unground configuration, compared with the close-to-zero conductance when grounding the edge. **f,** The same data as **(e)**, but magnify the y-axis. **g,** The same data as **(e)**, but plotted as $dV/dI$ versus $V_{bg}$. **h,i,** The same technique as above are used in Supplementary device #2 to validate the existence of edge states.

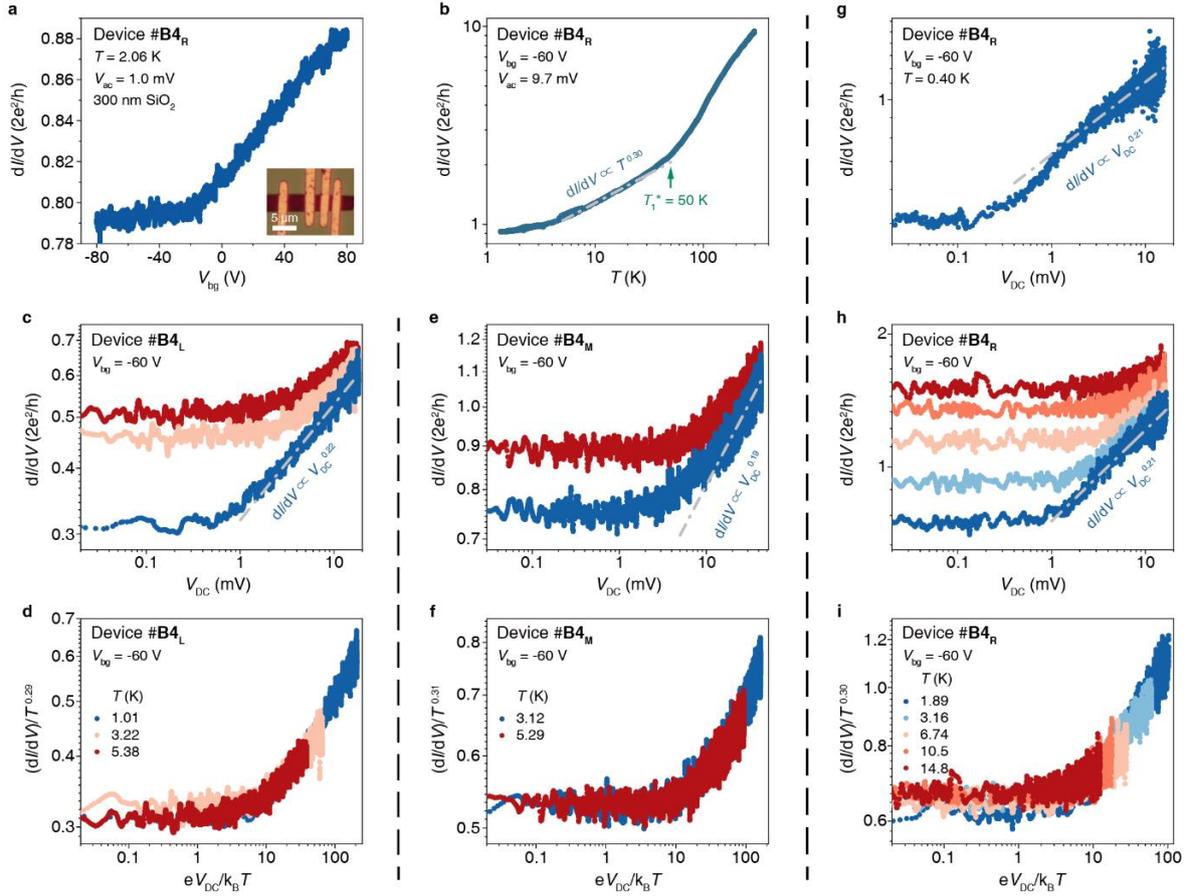

**Supplementary Fig. 8 Luttinger liquid behaviors in device #B4. a,** d$I$/d$V$ versus $V_{bg}$ at $T$ = 2.06 K measured with two-terminal configuration using the right two leads in device #B4 with 300 nm SiO$_2$ bottom gate. Inset: Optical image of device #B4. **b,** d$I$/d$V$ versus $T$ at $V_{bg}$ = -60 V measured with two-terminal configuration using the right two leads in device #B4. Note that the slight saturation below ~ 6 K is due to large a.c. voltage we applied. **c,d,** d$I$/d$V$ v.s. $V_{DC}$ (**c**) and (d$I$/d$V$)/$T^{0.29}$ v.s. e$V_{DC}$/k$_B T$ (**d**) measured at $V_{bg}$ = -60 V with two-terminal configuration using the left two leads in device #B4. **e,f,** d$I$/d$V$ v.s. $V_{DC}$ (**e**) and (d$I$/d$V$)/$T^{0.31}$ v.s. e$V_{DC}$/k$_B T$ (**f**) measured at $V_{bg}$ = -60 V with two-terminal configuration using the two middle leads in device #B4. **g,** d$I$/d$V$ v.s. $V_{DC}$ measured at $V_{bg}$ = -60 V and $T$ = 0.40 K with two-terminal configuration using the right two leads in device #B4. It exhibits sight edge gap behavior at $T$ = 0.40 K which is not observable above 1.89 K. **h,i,** d$I$/d$V$ v.s. $V_{DC}$ (**h**) and (d$I$/d$V$)/$T^{0.30}$ v.s. e$V_{DC}$/k$_B T$ (**i**) measured at $V_{bg}$ = -60 V with two-terminal configuration using the right two leads in device #B4. The gray dot dash lines show power-law behaviors. The universal scaling behaviors and consistent power law exponents α indicate that Littinger liquid behavior is the intrinsic property of Ta$_2$Pd$_3$Te$_5$.

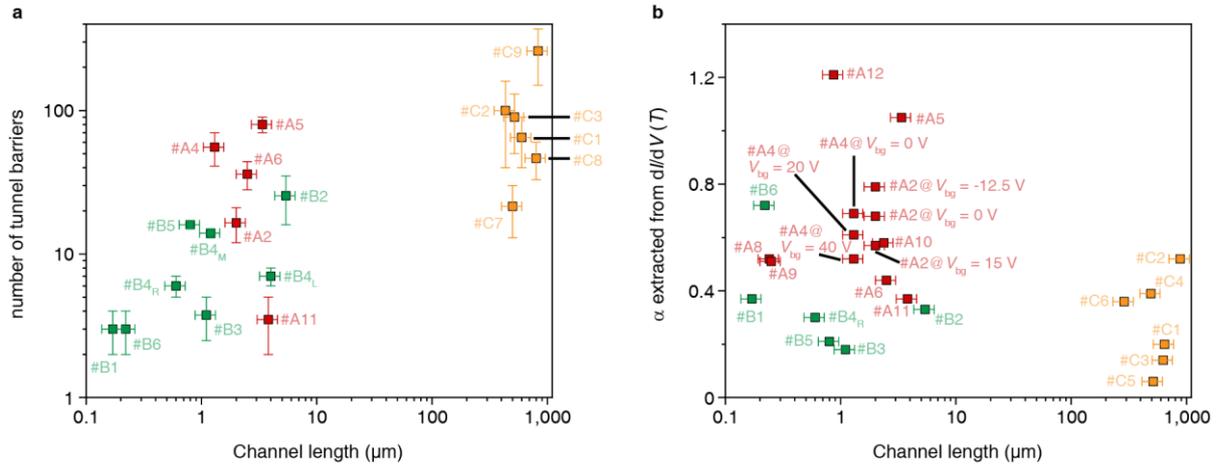

**Supplementary Fig. 9 Statistical result of the relation between number of tunnel barriers (a)/power exponents (b) and channel length of the devices/samples.** The error bars in *x*-axis reflect the uncertainty from channel length measurement and error bars in *y*-axis in **(a)** reflect the uncertainty of the extracted numbers. Note that the d$I$/d$V$ ($V_{DC}$) curves used to extract the number of tunnel barriers of bulk samples in **(a)** and d$I$/d$V$ ($T$) curves used to extract $\alpha$ of bulk samples in **(b)** are measured in different cryostats. The leads spacing are different between two measurements. So, the channel length of Bulk #C1-C3 are different between **(a)** and **(b)**.

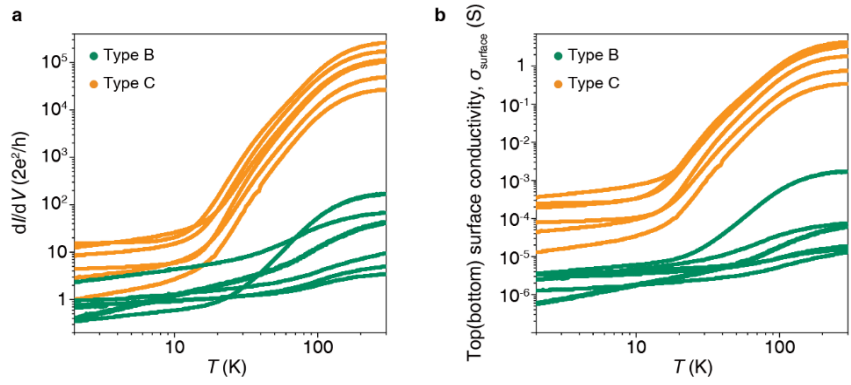

**Supplementary Fig. 10 Temperature dependence d$I$/d$V$ (a) and top(bottom) surface conductivity $\sigma_{\text{surface}}$ (b, defined as conductance / width × length) for type B (green line) and C (orange line) samples.** The curves in **(b)** diverge a lot in Luttinger liquid behavior region, which can exclude that Luttinger liquid behavior arises from top and/or bottom surface of $Ta_2Pd_3Te_5$ thin film devices/bulk samples.

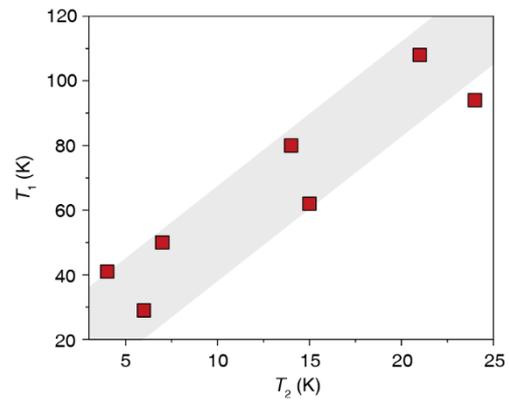

**Supplementary Fig. 11 $T_1$ versus $T_2$ for type A samples.** It shows approximate linear behavior. Uncertainties in both *x* and *y*-axis are smaller than point size.

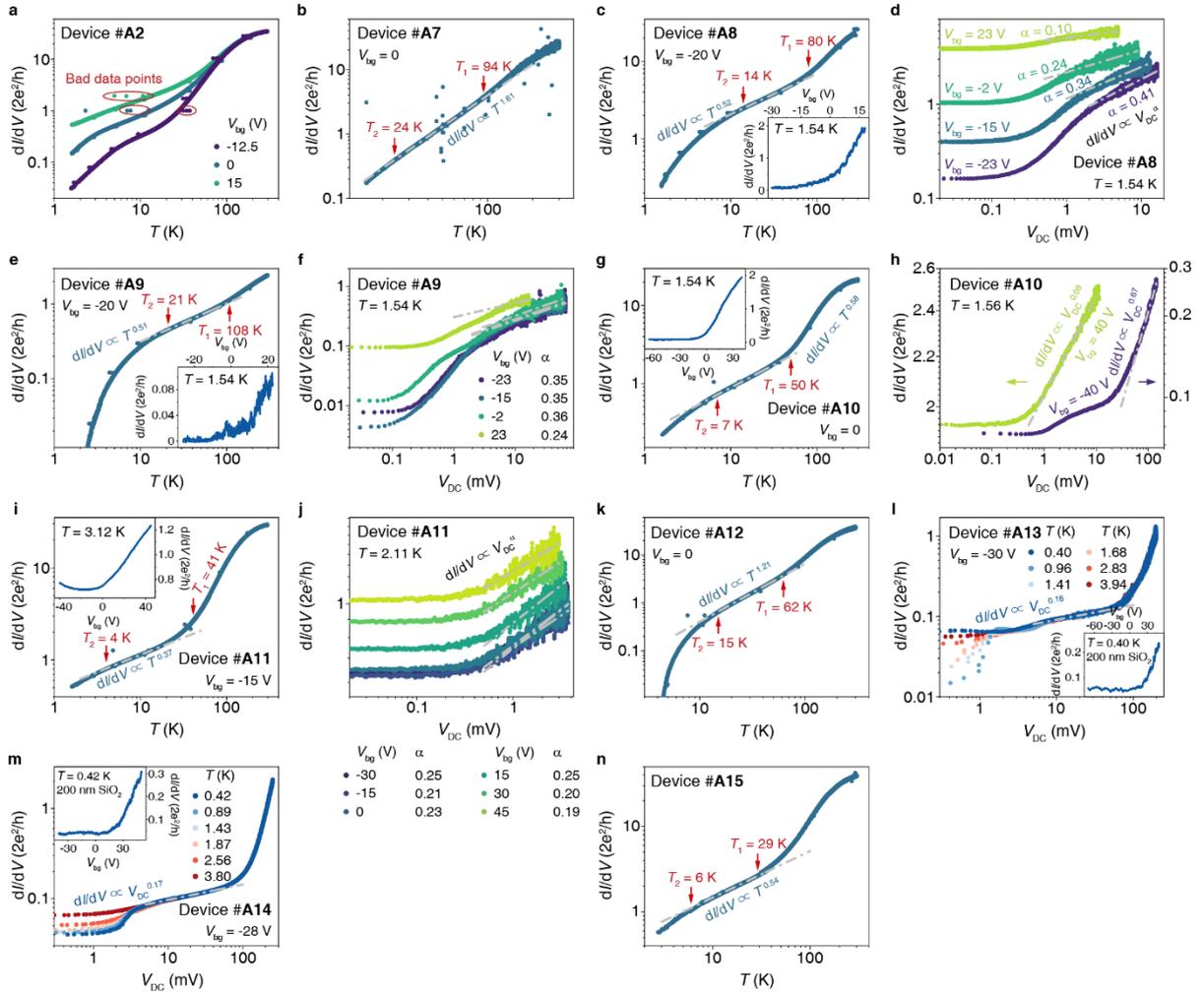

**Supplementary Fig. 12 Original data for type A devices. a,b,** Original temperature dependence d$I$/d$V$ data of device #A2 **(a)** and device #A7 **(b)**. (Bad data points due to thermometer range switches are removed in d$I$/d$V$ ($T$) curves in main figures and raw data are shown in Supplementary Figures. These bad data points don't affect the trend of d$I$/d$V$ ($T$) curves.) **c,d,** d$I$/d$V$ versus $T$ **(c)** and $V_{DC}$ **(d)** of device #A8 with the inset shows gate tunable d$I$/d$V$ at $T$ = 1.54 K. **e,f,** d$I$/d$V$ versus $T$ **(e)** and $V_{DC}$ **(f)** of device #A9 with the inset shows gate tunable d$I$/d$V$ at $T$ = 1.54 K. **g,h,** d$I$/d$V$ versus $T$ **(g)** and $V_{DC}$ **(h)** of device #A10 with the inset shows gate tunable d$I$/d$V$ at $T$ = 1.54 K. **i,j,** d$I$/d$V$ versus $T$ **(i)** and $V_{DC}$ **(j)** of device #A11 with the inset shows gate tunable d$I$/d$V$ at $T$ = 3.12 K. **k,** Temperature dependence d$I$/d$V$ of device #A12. **l,** d$I$/d$V$ versus $V_{DC}$ of device #A13 with the inset shows gate tunable d$I$/d$V$ at $T$ = 0.40 K. **m,** d$I$/d$V$ versus $V_{DC}$ of device #A14 with the inset shows gate tunable d$I$/d$V$ at $T$ = 0.42 K. **n,** Temperature dependence d$I$/d$V$ of device #A15. The gray dot dash lines show power-law behaviors.

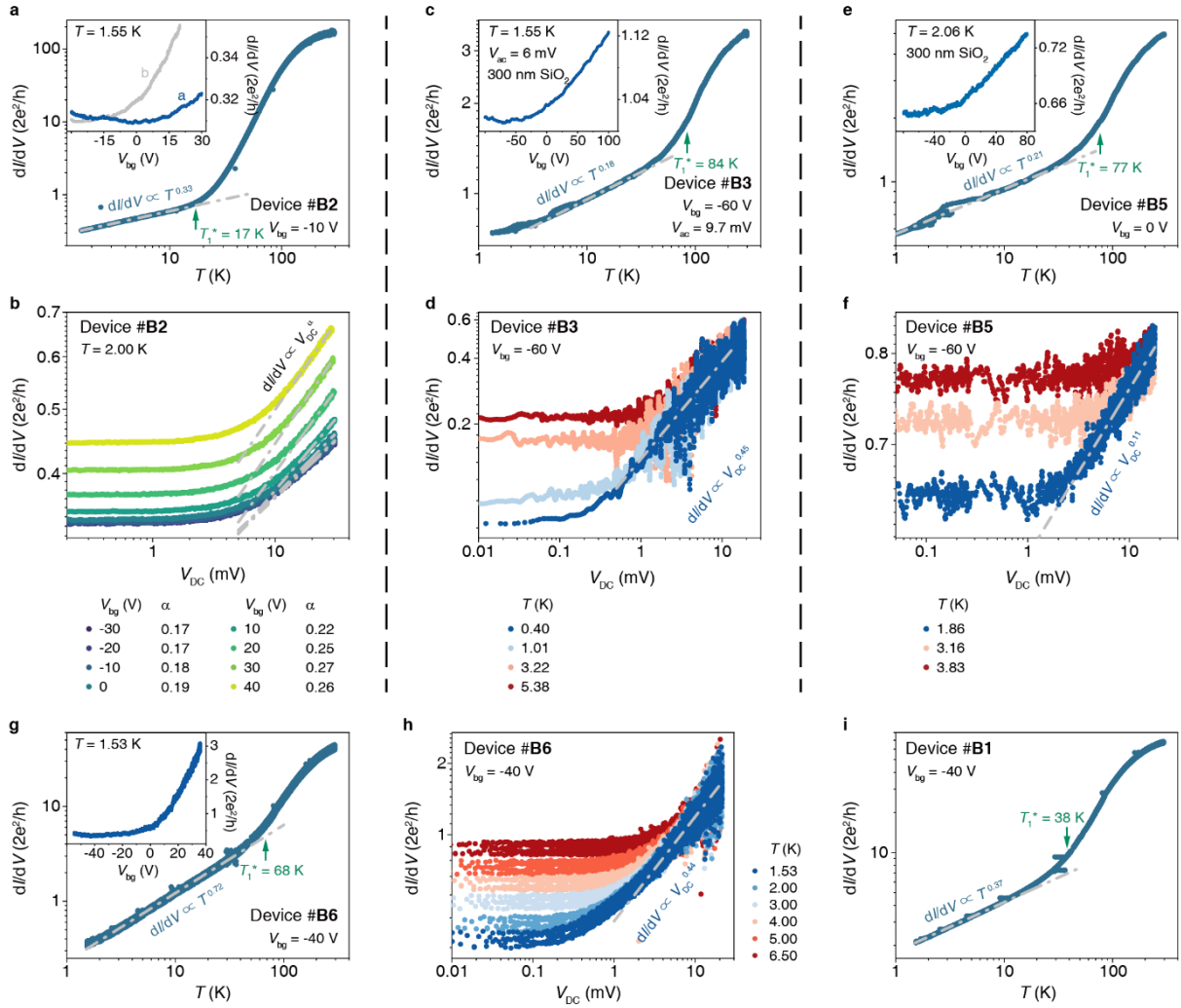

**Supplementary Fig. 13 Original data for type B devices. a,b,** d$I$/d$V$ versus $T$ **(a)** and $V_{DC}$ **(b)** of device #B2 with the inset shows gate tunable d$I$/d$V$ at $T$ = 1.55 K. Note that the CNP changed after several times of cooldown, then the blue line 'a' in the inset is for **(a)** and gray line 'b' for **(b)**. **c,d,** d$I$/d$V$ versus $T$ **(c)** and $V_{DC}$ **(d)** of device #B3 with the inset shows gate tunable d$I$/d$V$ at $T$ = 1.55 K. Note that the slight saturation in the lowest temperature region in d$I$/d$V$ versus $T$ is due to large a.c. voltage we applied. **e,f,** d$I$/d$V$ versus $T$ **(e)** and $V_{DC}$ **(f)** of device #B5 with the inset shows gate tunable d$I$/d$V$ at $T$ = 2.06 K. **g,h,** d$I$/d$V$ versus $T$ **(g)** and $V_{DC}$ **(h)** of device #B6 with the inset shows gate tunable d$I$/d$V$ at $T$ = 1.53 K. **i,** Original temperature dependence d$I$/d$V$ data of device #B1.

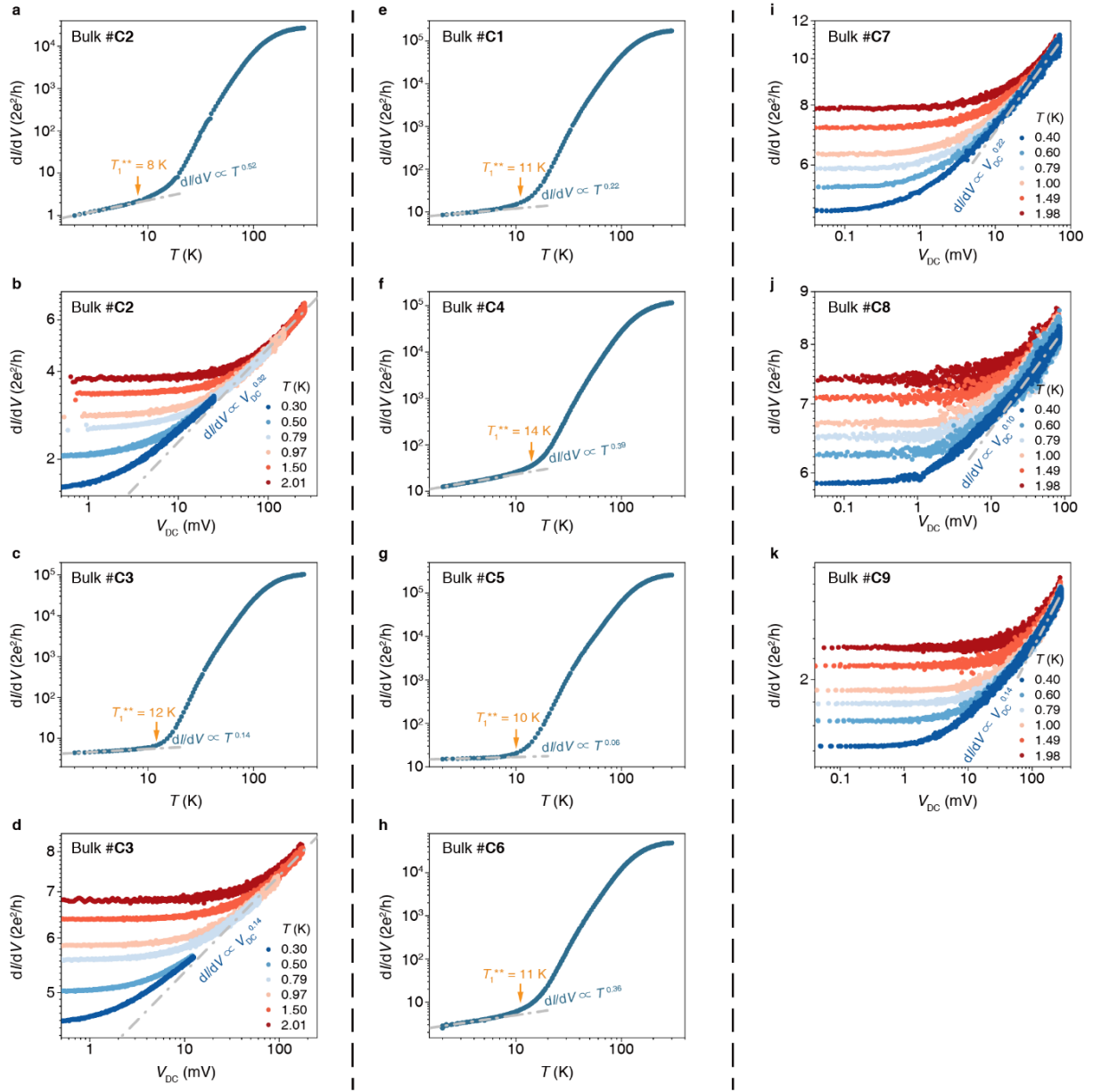

**Supplementary Fig. 14 Original data for type C bulk samples. a,b,** d$I$/d$V$ versus $T$ **(a)** and $V_{DC}$ **(b)** of bulk sample #C2. **c,d,** d$I$/d$V$ versus $T$ **(c)** and $V_{DC}$ **(d)** of bulk sample #C3. **e-h,** Temperature dependence d$I$/d$V$ of bulk samples #C1 **(e)**, #C4 **(f)**, #C5 **(g)**, and #C6 **(h)**. Note that due to different leads spacing in two measurements, the values of d$I$/d$V$ are slightly different between **(e)** and Fig. 5f. **i-k,** d$I$/d$V$ versus $V_{DC}$ of bulk samples #C7 **(i)**, #C8 **(j)**, and #C9 **(k)**.